\documentclass[twocolumn]{aastex631}
\UseRawInputEncoding
\usepackage{mathrsfs}
\usepackage{subfigure}
\usepackage{lineno}
\usepackage{natbib}
\usepackage{color}
\usepackage{ulem}
\usepackage{xcolor}

\def\degree{${}^{\circ}$}

\shortauthors{Chen et al.}

\begin{document}
\title{Estimating Bolometric Luminosities of Type 1 Quasars with Self-Organizing Maps}

\author[0000-0002-4765-1500]{Jie Chen}
\affiliation{Department of Astronomy, School of Physics, Peking University, Beijing 100871, China}
\affiliation{Kavli Institute for Astronomy and Astrophysics, Peking University, Beijing 100871, China}

\author[0000-0003-4176-6486]{Linhua Jiang}
\affiliation{Department of Astronomy, School of Physics, Peking University, Beijing 100871, China}
\affiliation{Kavli Institute for Astronomy and Astrophysics, Peking University, Beijing 100871, China}

\author[0000-0002-1234-552X]{Shengxiu Sun}
\affiliation{Department of Astronomy, School of Physics, Peking University, Beijing 100871, China}
\affiliation{Kavli Institute for Astronomy and Astrophysics, Peking University, Beijing 100871, China}

\author[0000-0002-2420-5022]{Zijian Zhang}
\affiliation{Department of Astronomy, School of Physics, Peking University, Beijing 100871, China}
\affiliation{Kavli Institute for Astronomy and Astrophysics, Peking University, Beijing 100871, China}

\author[0000-0002-0771-2153]{Mouyuan Sun}
\affiliation{Department of Astronomy, Xiamen University, Xiamen, 
Fujian 361005, China}

\begin{abstract}

We present a new method to calculate bolometric luminosities for unobscured, type 1 quasars with multi-band photometric data. Bolometric luminosity is a fundamental property to understand quasars and it is commonly estimated from monochromatic luminosities using bolometric corrections that often neglect quasar SED diversity. We take advantage of the fact that most quasars now have multi-band observations from UV to mid-IR, and construct SEDs for a well-defined sample of SDSS quasars at $0.5\leq z\leq 2$. Based on this fiducial sample, we explore quasar SEDs, their diversity, and their relations with bolometric luminosities. We then use unsupervised neural network self-organizing maps (SOM) to describe the SED diversity and compute the bolometric luminosities with a fully-trained SOM model. This method reduces systematical uncertainties compared to the traditional method. In addition, we update the multi-linear regression relations between bolometric luminosity and monochromatic luminosities at restframe 1450\,\r{A}, 3000\,\r{A}, and 5100\,\r{A}. Our method is applicable to large quasar samples with a wide range of luminosity and redshift. We have applied it to the SDSS DR16 quasars. We have also made our code publicly available.
 
\end{abstract}


\section{Introduction} \label{sec:intro}

Quasars show strong radiation across the entire electromagnetic spectrum, powered by the accretion of central supermassive black holes (SMBHs). Their spectral energy distributions (SEDs) reveal the physical properties of the central black hole, accretion disc, dust torus, and possible radio emission. The bolometric luminosity ($L_\mathrm{bol}$) of a quasar is defined as the total energy per second over all wavelengths in all directions, and is often computed as the integrated emission of its SED. It is a crucial parameter to understand the SMBH accretion \citep[e.g.,][]{1983ApJ...269..352S,2003ApJ...598..886U,2003ApJ...582..559V,2007ApJ...654..731H}. The bolometric luminosity also represents the mass accretion rate of a SMBH by $L_\mathrm{bol}=\eta \dot M c^2$, where $\dot M$ is the absolute mass accretion rate and $\eta$ is the accretion efficiency. Combined with the black hole mass ($M_{\rm{BH}}$), we can obtain the black hole accretion rate using $\lambda_\mathrm{Edd}\equiv L_\mathrm{bol}/L_\mathrm{Edd}$, where $L_\mathrm{Edd}$ is the Eddington luminosity. The Eddington ratio is a key parameter that determines the accretion disk’s geometric structure and dynamical properties \citep{2014ARA&A..52..529Y}.

Measuring bolometric luminosity presents significant challenges. We summarize the procedure and associated challenges below. 
\begin{enumerate}

\item First of all, measuring accurate bolometric luminosity requires substantial investment in telescope time to conduct observations across the full wavelength SED. 

\item It is controversial whether or how to include radiation in the mid-IR and longer wavelength. Some studies include mid-IR emission from the dust torus \citep[e.g.,][]{1994ApJS...95....1E,2006ApJS..166..470R}. Others argue that including mid-IR emission would overestimate the total luminosity because it is the reprocessed radiation of the UV/optical photons that is already considered in the SED \citep[e.g.,][]{2004MNRAS.351..169M,2012MNRAS.422..478R,Rosario2013,2016ApJ...833...98C}. In addition, it is unclear how to include radiation in the longer wavelengths that is presumably dominated by the host galaxy.

\item It is difficult to remove the contamination from the host galaxy that contributes more than 30$\%$ of the total flux at 5100\,\r{A} in low redshift  (z$<$0.8) quasars or AGNs \citep[e.g.,][]{2011ApJS..194...45S,2023MNRAS.521L..11J,2024arXiv240617598R}. The host fraction depends on luminosity. There are different approaches to estimate the host galaxy contribution. For example, some studies performed a SED fitting to decompose a quasar to an AGN component and a galaxy component \citep[e.g.,][]{2012MNRAS.425..623L,Lyu2017,2020A&A...636A..73D,2023ApJ...954..156K}. Others used an elliptical galaxy template and one band host galaxy fraction to correct the host galaxy contamination \citep[e.g.,][]{2006ApJS..166..470R,2012MNRAS.422..478R,2013ApJS..206....4K}. Some studies have shown that most low-redshift quasars are in star-forming galaxies \citep[e.g.,][]{Rosario2013,Shangguan2020, Zhuang2021}.

\item It is challenging to obtain the extreme-UV (EUV) radiation of a quasar. The EUV radiation contributes a large portion of a quasar SED, but it cannot be directly observed because of the absorption by the galactic and extragalactic hydrogen atoms \citep{Lynds1971}. Different studies in the literature often used different models with different EUV SEDs that can affect the bolometric luminosity measurement \citep{2012MNRAS.422..478R,2013ApJS..206....4K}. 

\item We often assume that the quasar radiation is isotropic when we calculate the bolometric luminosity. This is apparently not correct. For example, the mid-IR emission from the dust torus is not isotropic, and the optical-UV emission from the accretion disk is highly anisotropic \citep{2016MNRAS.458.2288S}. Hence, an average viewing angle is needed to adjust the bolometric luminosity \citep[e.g.,][]{2001ApJ...559..680H,2010Nemmen,2012MNRAS.422..478R}. 

\end{enumerate}

Computing the isotropic bolometric luminosity of a quasar with a full wavelength SED is relatively straightforward. It is challenging to obtain bolometric luminosities for large numbers of quasars with photometry in a few bands. Bolometric correction (BC) is a widely used method to estimate bolometric luminosity based on a known monochromatic luminosity. The BC method depends on empirical correlations between the bolometric luminosity and X-ray luminosity \citep[e.g.,][]{2017ApJ...844...10B,2020A&A...636A..73D} or continuum luminosities in the UV and optical bands \citep[e.g.,][]{2006ApJS..166..470R,2013ApJS..206....4K,2012MNRAS.422..478R,2020A&A...636A..73D}. Some studies use the IR luminosity only to compute the bolometric luminosity, assuming that the radiation is almost isotropic \citep{2012MNRAS.426.2677R,2023ApJ...954..156K}. In addition, BCs are shown to be luminosity-dependent \citep[e.g.,][]{2004MNRAS.351..169M,2013ApJS..206....4K,2020A&A...636A..73D} or Eddington ratio-dependent \citep[e.g.,][]{2007MNRAS.381.1235V,2012MNRAS.425..907J}. On the other hand, the diverse SEDs of quasars make it inappropriate to use a simple template to estimate bolometric luminosity \citep{2006ApJS..166..470R,2023ApJ...957...19A}. 

Many quasars now benefit from multi-wavelength observations provided by surveys such as Sloan Digital Sky Survey \citep[SDSS;][]{York2000}, the Galaxy Evolution Explorer \citep[GALEX;][]{2005ApJ...619L...1M}, the UK Infra-Red Telescope (UKIRT) Hemisphere Survey \citep[UHS;][]{2018Dye, Bruursema2023USNO}, the Wide-Field Infrared Survey Explorer \citep[WISE;][]{2010AJ....140.1868W}. These datasets span a wide range of the electromagnetic spectrum, offering a comprehensive view of quasar SEDs. Leveraging this rich multi-band data enables reliable calculation of bolometric luminosities. In principle, incorporating multi-wavelength information would significantly alleviate biases introduced by the luminosity dependency and SED diversity \citep{Su2025}. Despite this fact, current studies still largely rely on the usage of a single monochromatic luminosity and the BC method \citep[e.g., SDSS DR16Q in][]{2022ApJS..263...42W}. 

In this work, we define and calculate isotropic bolometric luminosity under the assumption of an unobscured quasar, $L_{\rm{bol}}=\int_0^\infty 4\pi D_L^2 F_\nu d\nu$, where $D_L$ and $F_\nu$ are the luminosity distance and rest-frame monochromatic flux. The true anisotropic bolometric luminosity is $f L_{\rm{bol}}$, where $f$ is a correction factor. For example,  \citet{2010Nemmen} found that $f$ is roughly 0.75. Unless stated otherwise, $L_{\rm{bol}}$ denotes isotropic luminosity.

We are motivated to improve the isotropic luminosity $L_{\rm{bol}}$ computation based on multi-band observations rather than a composite SED or single monochromatic luminosity. Quasars that we consider here are typical unobscured type 1 quasars. We mainly use the SDSS-IV quasars from Data Release 16 \citep[DR16Q;][]{2020ApJS..250....8L}. We collect multi-band observations from the mid-IR to UV for each quasar and use the machine learning algorithm Self-Organizing Map \citep[SOM;][]{KOHONEN82} to calculate $L_{\rm{bol}}$. This method accelerates the calculation of bolometric luminosities and enhances the accuracy by accounting for the correlations between photometric bands. Utilizing multi-band data reduces the systematic uncertainty of bolometric luminosity relative to the single-band BC method. We make the code publicly available. 

This paper is formatted as follows. Section~\ref{Sec:Data} introduces our data and quasar sample selection. Section~\ref{Sec:SED} describes the SED construction and the bolometric luminosity calculations. In section~\ref{Sec:SOM}, we apply the SOM algorithm to our sample to computed bolometric luminosities. Section~\ref{Sec:Summary} summarizes the paper. We adopt a cosmology with $H_0 = 70$\,km\,s$^{-1}$\,Mpc\,$^{-1}$, $\Omega_{\rm{M}} = 0.3$ and $\Omega_\Lambda = 0.7$. The Eddington luminosity is $L_{\mathrm{Edd}}=1.3\times 10^{38}\ (M_{\mathrm{BH}}/M_{\odot})\ \mathrm{erg\ s^{-1}}$.

\section{DATA AND QUASAR SAMPLE}\label{Sec:Data}

We use the SDSS-IV quasar catalog from Data Release 16 \citep[DR16Q;][]{2020ApJS..250....8L} that contains 750,414 spectroscopically confirmed quasars. Each quasar has photometric data in the five SDSS optical bandpasses \citep[$ugriz$;][]{1996AJ....111.1748F}. We use 197,588 relatively bright quasars with $r\leq 20.4\, \rm{mag}$ in the redshift range of $0.5\leq z\leq 2$, after correction for the Galactic extinction. In this redshift range, the $u$ and NUV bands cover the peak of the quasar SED in the rest-frame far-UV. We restrict our study to unobscured, type 1 quasars, and quasars are required to have broad Mg II emission lines (2798\AA) in the SDSS spectra based on the catalog of \cite{2022ApJS..263...42W}. For each quasar, we retrieve its $E(B-V)$ from the \cite{1998ApJ...500..525S} dust map and correct for the Galactic extinction using the extinction coefficients provided by \cite{2011ApJ...737..103S}. We keep quasars with a signal-to-noise ratio (S/N) greater than 5 in all five SDSS bands, and the total quasar number is 193,822.

\begin{figure}
\centering
\includegraphics[width=0.5\textwidth]{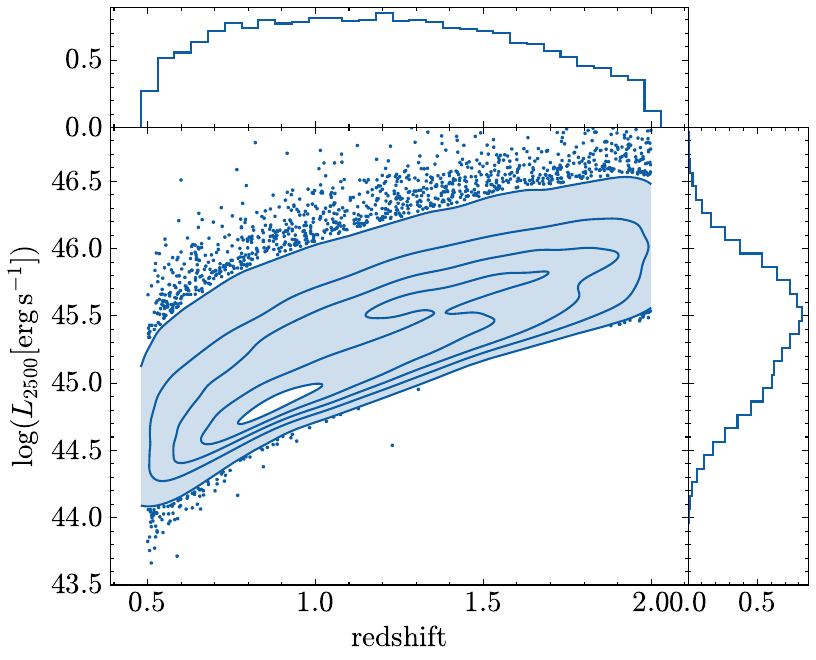}
\caption{Redshift and $L_{2500}$ distributions of the quasars in our sample. The quasars are represented by the blue contours. The contours indicate 0.005, 0.25, 0.5, 0.75, and 0.95 of the normalized distributions. The histograms of redshift and luminosity are normalized by the total number of the quasars.}
\label{fig:L-z}
\end{figure}

In order to obtain the photometry of the quasars in other bands, we use the SDSS optical positions to crossmatch the quasars with the data from the Galaxy Evolution Explorer \citep[GALEX;][]{2005ApJ...619L...1M}, the UKIRT Infrared Deep Sky Survey \citep[UKIDSS;][]{2007MNRAS.379.1599L}, the UK Infra-Red Telescope (UKIRT) Hemisphere Survey \citep[UHS;][]{2018Dye,Bruursema2023USNO}, Vista Hemisphere Survey \citep[VHS;][]{2013McMahon}, the Wide-Field Infrared Survey Explorer \citep[WISE;][]{2010AJ....140.1868W}. The final sample consists of 79,163 quasars (see details in the following subsections). 
The Mg II FWHM values of this sample are greater than 4000 $\rm{km}\,s^{-1}$. Figure~\ref{fig:L-z} shows the distributions of the redshifts and monochromatic luminosities $L_{2500}$ (i.e., $\lambda L_{\lambda}$ at 2500\,\r{A}) for the sample. $L_{2500}$ is calculated by an interpolation from the closest filters. Figure~\ref{fig:SED} shows the multi-band SEDs of the quasars, and Table~\ref{tab:quasar_catalog} in Appendix~\ref{appendixA} provides the multi-band photometry of the sample.

Our quasars are luminous and at redshift about 0.5 to 2. They are point-like sources in ground-based images or GALEX and WISE images. We use total magnitudes or total flux to describe them in each band. For SDSS, we use PSFMAG. For other data, we adopt aperture magnitudes with aperture corrections applied (see details in the following subsections).

\begin{figure*}
\centering
\includegraphics[width=0.7\textwidth]{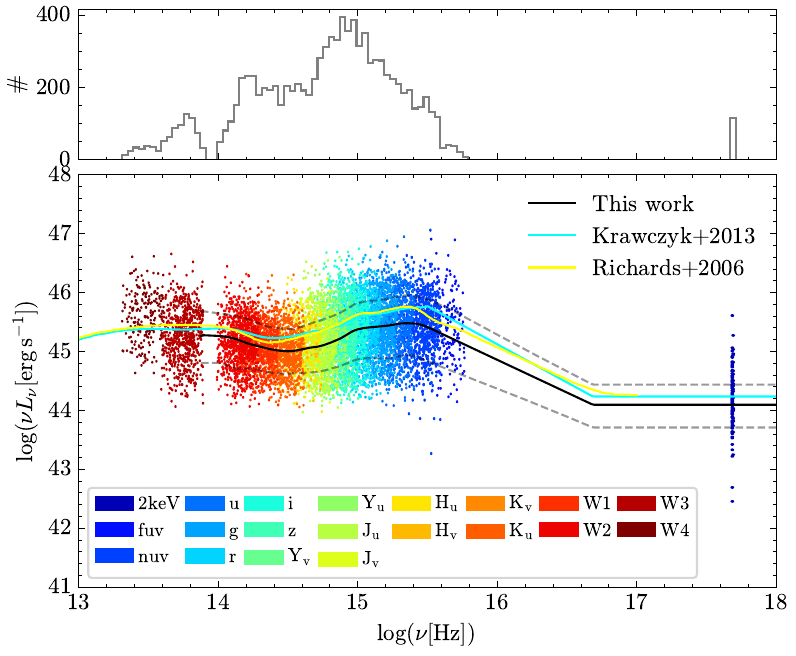}
\caption{Multi-band SEDs of the quasars in our sample. The upper panel shows the number of SEDs averaged at each frequency. In the lower panel, the solid curve represents the median SED of all quasars in the sample and the dashed curves represent its $1\sigma$ ranges. The data points are photometric data of randomly selected 800 quasars.}
\label{fig:SED}
\end{figure*}

\subsection{Near-IR Data}

In the near-IR band, we use the public catalog from UKIDSSDR11PLUS, UHS DR2, and VHS DR6. We crossmatch SDSS DR16Q with these surveys using a radius of $1\arcsec$. The survey depths of the UKIDSS Large Area Survey (ULAS) and UKIDSS Galactic Cluster Survey (GCS) in the Y, J, H, and K bands are 20.2 (20.1), 19.6, 18.8, and 18.2 mag (Vega). The UHS is an ongoing UKIRT program that surveys the Northern Hemisphere in the Y, J, H, and K bands. For now, the J and K bands have been publicly available, with a median $(5\sigma)$ point source sensitivity of 19.6 mag in J and 18.4 mag in K (Vega). 
Here magnitudes are total magnitudes corrected from aperture magnitudes APERMAG3 within a $2\arcsec$ aperture in diameter.
The VHS survey observed the southern sky hemisphere in the Y, J, H, and Ks bands, with the depths $(5\sigma)$ of 20.6, 20.2, 19.2, and 18.2 Vega mag in the four bands. For the overlap regions of ULAS, GCS, and UHS, we use the average photometry from two catalogs. We require that quasars in the final sample should be detected with S/N greater than 5 in at least two near-IR bands. We keep the sources with the flag ERRBITS=0. In total, we have 128,489 matches from the near-IR surveys.

\subsection{Mid-IR Data}

We match 128,498 SDSS quasars (with a match length of $1\arcsec$) to the AllWISE Data Release \citep{2014yCat.2328....0C} to extend their SEDs into the mid-IR range. The WISE bandpasses are generally referred to as W1 through W4, with effective observed frame wavelengths of 3.36, 4.61, 11.82, and 22.13 $\mu$m, respectively. We use w1mpro to w4mpro from ALLWISE, a profile-fit photometry magnitude. We require that all matched quasars have S/N $\geq 5$ in both W1 and W2. If matched quasars have S/N $\geq 3$ in W3 and W4, we also include W3 and W4 in their SEDs. We keep the sources with flag cc\_flag=0 to exclude the contamination and confusion. In total, we have 123,489 matches with AllWISE.

\subsection{UV Data}

In the UV, we use data from two GALEX bands, FUV (1350-1750\,\r{A}) and NUV (1750-2750\,\r{A}). We start with the sample of 123,498 quasars that satisfy the above IR requirements. We keep quasars only if they have detections in both NUV and FUV bands to avoid an extra scatter in the UV SED. We further require that the positions of the quasars are within the central 0.5-degree radius of the field-of-view (FOV) for any GALEX tiles \citep{2014Bianchi,2017Bianchi} to avoid poor GALEX photometry/astrometry. We then search the GALEX GR6Plus7 catalog for quasar counterparts within a matching radius of $2.6\arcsec$ \citep{2007Trammell}. Following a conservative recommendation, sources with artifact flags of 4 or 2 from \cite{2017Bianchi} are removed. For sources detected in multiple tiles, we follow \cite{2023NatAs...7.1506C} and adopt the one with the longest FUV exposure time. We use MAG\_AUTO from GALEX, a Kron elliptical aperture magnitude.

In order to correct the Galactic extinction in the GALEX bands, we adopt $A_{\lambda}/E(B-V)_{\rm{SFD}}$ = 6.783 and 6.620 for the GALEX FUV and NUV bandpasses, respectively. The correction factors are from \cite{2013MNRAS.430.2188Y} who used the \cite{1999PASP..111...63F} reddening law with $R_V$= 3.1. We require that the matched quasars should have S/N $\geq 3$ in NUV. Photometry with S/N $\geq 5$ in FUV is included in SED. 

The full width at half-maximum of the GALEX point spread function is about $5\arcsec$ \citep{2007ApJS..173..682M}, so blending and contamination from nearby bright sources are an issue. We check the SDSS images in the GALEX (quasar) positions with a radius of $2.6\arcsec$. We exclude 138 quasars that have at least one neighbor brighter than $m_{\rm quasar}+1$ mag in the $u$ band.

Our quasar sample spans an FUV brightness range from 17 to 24 mag, which is in the faint regime of the FUV flux nonlinearity calibration \citep{2014MNRAS.438.3111C,2019MNRAS.489.5046W,2023MNRAS.523.4067W}. The calibration is based on several white dwarfs to calibrate the nonlinearity for sources brighter than 16 mag. Since our quasars are fainter than the calibration threshold, we assume that no additional correction is required. Finally, we obtain 79,163 matches with GALEX.

\subsection{X-Ray Data}

We crossmatch SDSS DR16Q with XMM-Newton Serendipitous Source Catalog \citep[4XMM DR13;][]{2020A&A...641A.136W,2020A&A...641A.137T}, Chandra Source Catalog \citep[CSC v2.1;][]{2024ApJS..274...22E}, and SRG/eROSITA all-sky survey \citep[eRASS DR1;][]{2024A&A...682A..34M}. For quasars detected in both soft (0.5-1.0 keV for XMM-Newton, 0.2-0.5 keV for eROSITA, and 0.5-1.2 keV for Chandra) and hard (2.0-5.0 keV for XMM-Newton, 1.0-2.0 keV for eROSITA, and 0.5-1.2 keV for Chandra) X-ray bands, we estimate the intrinsic absorption column density using their soft-to-hard band ratios, assuming an intrinsic power-law model modified by the Galactic absorption \citep[][]{2016A&A...594A.116H} and intrinsic absorption. The intrinsic power-law photon index $\Gamma$ is assumed to be 1.7, which is typical for distant X-ray AGNs \citep[e.g.,][]{2015A&A...577A.121R}. We then calculate the absorption-corrected 2 keV flux. For quasars that are not detected in the hard or soft bands, we assume $\Gamma = 1.7$ and estimate their 2 keV flux using the flux from the nearest detected bands. All X-ray luminosities are corrected for the Galactic absorption. In total, we obtain 11,825 matches with these X-ray surveys in 79,163 quasars.

Due to the limited sky coverage of sensitive X-ray observations, the number of X-ray-detected quasars is significantly lower than those detected in the optical and IR bands. For quasars without X-ray detections (85\%), we estimate their X-ray flux using the correlation defined by the $L_{\rm{UV}}-L_{\rm{X}}$ relation based on the luminosities at 2500\,\r{A} and 2 keV. \cite{2006AJ....131.2826S} used 333 quasars with $z\leq 6$ and $\log (\lambda L_{\lambda}/\rm{erg\,s^{-1}})>42$ at 2500\,\r{A} and find the following $L_{\rm{UV}}-L_{\rm{X}}$ relation
\begin{equation}
    \log (L_{2 \mathrm{keV}})=(0.721 \pm 0.011) \log (L_{2500})+(4.531 \pm 0.688).
\end{equation}
We assume an X-ray energy spectral index $\alpha_x=-1$ (photon index $\Gamma=2$) between 0.2 keV and 10 keV \citep[e.g.,][]{2000ApJ...531...52G}.

\section{Construction of SEDs and calculation of bolometric luminosities} 
\label{Sec:SED}

In this section, we first introduce the corrections of the Galactic extinction and the intergalactic medium \citep[IGM;][]{2009ApJ...705L.113P} absorption, and construct quasar SEDs. We then calculate bolometric luminosities and estimate bolometric corrections. Finally, we apply multi-band linear regression to improve the accuracy of the bolometric luminosity calculation.

\subsection{IGM Absorption Correction}
\label{subsec:IGM}

\begin{figure}
\includegraphics[width=0.5\textwidth]{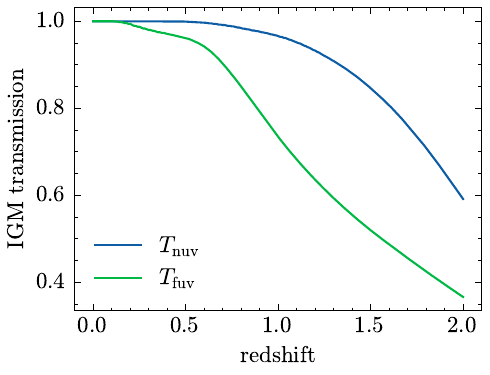}for a power law spectrum with an index of --0.44 as a function of redshift in the FUV and NUV bands.
\label{fig:IGM}
\end{figure}

The EUV ($\lambda_{\rm{rest}} < 1216$\,\r{A}) part of the quasar spectrum is significantly attenuated by the IGM. To statistically correct the IGM absorption, we model the IGM transmission as a result of absorption by both Lyman continuum (LyC) and Lyman series lines (LyL) following \cite{2023NatAs...7.1506C}. The total IGM transmission is $T_{\lambda} = \exp(-\tau_{\rm{eff}})$, where the total effective optical depth, $\tau_{\rm{eff}}= \tau_{\rm{eff}}^{\rm{LyC}} + \tau_{\rm{eff}}^{\rm{LyL}} $, with the effective LyC optical depth, $\tau_{\rm{eff}}^{\rm{LyC}}$, and the effective LyL optical depth, $\tau_{\rm{eff}}^{\rm{LyL}}$. The optical depth models the average attenuation of a source assuming Poisson-distributed neutral hydrogen (H\,I) clouds along the line of sight. Adopting the \cite{2020Faucher} distribution of absorbers with the H\,I column density of $ 12\leq \log N_{\rm{HI}} \leq 22$ and the Doppler broadening parameter of $b = 30\, \rm{km\, s^{-1}}$, we calculate the average IGM transmission curves for redshift from 0.5 to 2, with a step size of 0.01. We assume a power-law continuum with an index from the continuum fitting in \cite{2022ApJS..263...42W}, and compute its filter-weighted broadband mean IGM transmission to correct the photometry in the FUV and NUV bands. Figure~\ref{fig:IGM} shows an example of the mean broadband IGM transmission with an index of --0.44 for the two bands at each redshift \citep{2001AJ....122..549V}.

\subsection{Median SED and SED Diversity}

To illustrate the diversity of type 1 quasar SEDs, we first build a median quasar SED. To construct quasar SEDs, we convert their flux densities to luminosities and shift the bandpasses to the rest frame. All fluxes here are energy-weighted in SEDs. Figure~\ref{fig:SED} shows the SEDs of 800 randomly selected quasars from our samples. The Galactic extinction and IGM absorption have been corrected. We then create a data grid with a step size of 0.02 in log frequency, and linearly interpolate the quasars SEDs to match the grid. To generate the median energy distribution, we take a median value in $\log \nu L_{\nu}$ of all the data in each frequency bin, and it is shown as the black solid curve in Figure~\ref{fig:SED}. We use the 16th and 84th percentiles of the SEDs as the $1\sigma$ confidence level (i.e., the dashed line in the figure). The host galaxy contribution has not been removed in this step. For comparison, the mean quasar SEDs from \cite{2006ApJS..166..470R} and \cite{2013ApJS..206....4K} are shown in yellow and cyan, respectively. Our sample is slightly fainter, and thus our median SED is relatively low. We also try a brighter quasar sample with $r\leq19.1\,\rm{mag}$, and find that the median SED is consistent with those from \cite{2006ApJS..166..470R} and \cite{2013ApJS..206....4K}.

Figure~\ref{fig:SED-diversity} shows the diversity of the quasar SEDs in the UV to IR range. In this figure, we display 4000 randomly selected quasars from our sample and their SEDs are normalized at 5100 \AA. These 4000 UV-IR SEDs follow the same $L_{5100}$ distribution of the whole sample. The large scatter suggests a highly diverse nature of quasar SEDs that cannot be well explained by a single quasar template. Note that the $y$-axis is on a logarithmic scale that has reduced the scatter visually. The reason for the SED diversity is the combination of many factors, including host galaxy contributions, black hole masses, accretion rates, viewing angles, etc. The details are beyond the scope of this study. 

\subsection{Host Galaxy Correction}

Previous studies have found empirical anti-relations between the quasar host galaxy fraction and luminosity or redshift, but the scatters of these relations are typically very large \citep[e.g.,][]{2011ApJS..194...45S,2023MNRAS.521L..11J,2024arXiv240617598R}. Here we use the relation between the host fraction and luminosity at 5100\,\r{A} from \cite{2023MNRAS.521L..11J}. Some studies have shown that most low-redshift quasars are in star-forming galaxies \citep[e.g.,][]{Rosario2013, Shangguan2020, Zhuang2021}. Hence, we use the spiral galaxy template of \cite{2010Assef} to represent the host galaxy. Compared to elliptical galaxies, spiral galaxies have a larger short-wavelength contribution from the star-forming component. Removing galaxy components decreases the bolometric luminosity by an average of 0.05 dex. In the near future, the China Space Station Telescope and Euclid surveys will cover nearly half the sky in the wavelength range from NUV to near-IR, which will allow us to estimate host components for individual quasars by image decomposition. 

\subsection{Emission Line Correction}
We calculate the K-correction of the broadband photometry for individual sources based on the continuum fitting results  by \cite{2022ApJS..263...42W}. Due to the fixed wavelength coverage in the observed frame, the SDSS spectra do not cover some strong lines (e.g., Ly$\alpha$, C IV) for quasars in some particular redshift ranges, but they cover Mg II for all our quasars. Therefore, we use the empirical relation of equivalent width between Mg II and other lines (i.g.,  Ly$\alpha$, C IV, H$\beta$, and H$\alpha$ lines). We artificially add the lines in the extrapolated continuum and compute the corresponding K-correction from UV to near-IR. After the correction of the line emission, the bolometric luminosity decreases by about 0.02 dex on average.

\begin{figure}
\centering
\includegraphics[width=0.5\textwidth]{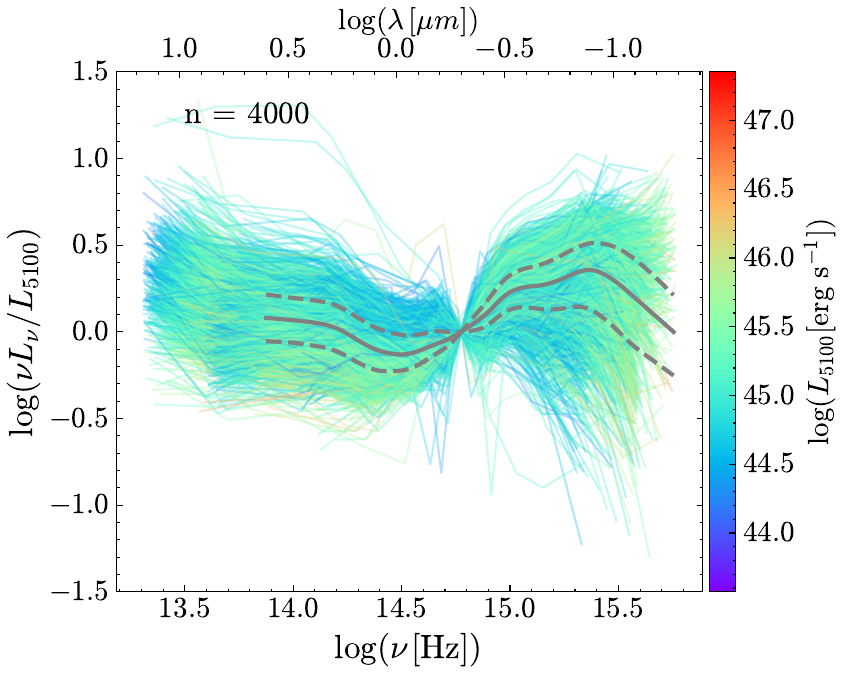}
\caption{Diversity of quasar SEDs. The color-coded curves represent 4000 randomly selected quasars from our sample. The SEDs have been normalized at 5100\,\r{A}. The gray solid curve is the median normalized SED with $1\sigma$ level indicated by the dashed curves.}
\label{fig:SED-diversity}
\end{figure}

\subsection{Bolometric Luminosity Measurement and Wavelength Range Selection}

\begin{figure*}
\centering
\includegraphics[width=1\textwidth]{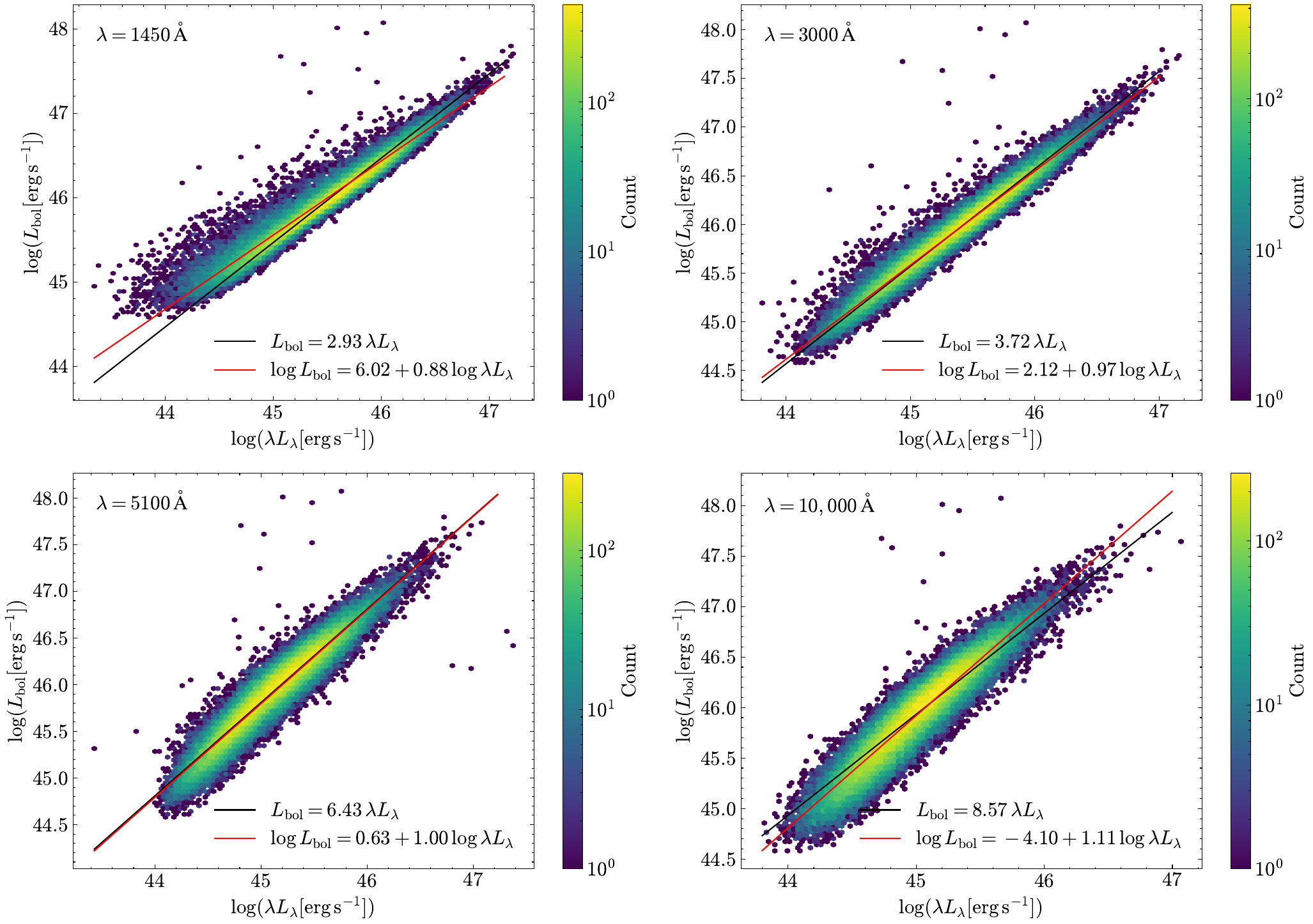}
\caption{Relation between bolometric luminosity and monochromatic luminosity at 1450\,\r{A}, 3000\,\r{A}, 5100\,\r{A}, and 10,000\,\r{A}. The black lines indicate the BC fitting and the red lines show the linear regression. }
\label{fig:Llambda-Lbol}
\end{figure*}

Bolometric luminosity is the integrated area under the SED curve. Its mathematical definition is as follows,
\begin{equation}
    L_{\mathrm{bol}}=\int_0^{\infty} L_\nu d \nu=\int_{-\infty}^{\infty} \ln (10) \nu L_\nu d \log (\nu).
\end{equation}
In practice, the observed wavelengths are limited. The choice of the integration range is an unclear issue as we discussed in the Introduction. Regarding the IR SED, some previous studies suggested that IR emission should not be included because it represents the reprocessed radiation of accretion disc photons by dusty torus \citep[e.g.,][]{2004MNRAS.351..169M}. There is a similar claim for the X-ray emission. Hard X-ray emission is generated through the inverse-Compton upscattering of disc photons by electrons in a hot corona. The reprocessing of these photons depends on the geometry of the corona \citep{2004ApJ...608...80S,2004ApJ...617..102S}. Therefore, the definition of the wavelength range for the bolometric luminosity computation is contingent upon the understanding of the structure surrounding the SMBH.

Our approach is to make the wavelength range flexible \cite[e.g.,][]{2013ApJS..206....4K}. We use an integration range of 4 \micron-10 keV (default range) in this paper. On the other hand, our code allows users to choose a different range for their own scientific purposes. About 20\% of our rest-frame SEDs do not reach 4 \micron\, because of the low detection rate in W3 and W4. We use \texttt{KNNImputer} \citep{Troyanskaya}, an imputation for completing missing values using k-Nearest Neighbors, to recover the missing SEDs with $n\_neighbors = 100 $ in our samples. The bolometric luminosities of full-coverage samples are calculated based on recovering SEDs. Conclusions for the paper remain quantitatively consistent when excluding these samples.

Calculating the uncertainty of bolometric luminosity is also challenging. Many sources of uncertainty exist, including but not limited to the SED interpolation, SED variation, host galaxy correction, orientation, etc. In this work, we do not consider these issues in detail except for the uncertainty from the interpolation. We emphasize that the uncertainty of bolometric luminosity is underestimated in this work and most other works.

\subsection{Isotropy and Viewing Angles}

One popular model to describe the SMBH gas accretion in AGNs is the geometrically thin but optically thick disk \citep[e.g.,][]{NT1973,SS1976}. A face-on accretion disc is usually UV-brighter than an edge-on disc, so bolometric corrections based on face-on observations may overestimate bolometric luminosities. Based on studies of the theoretically thin accretion disc models, \cite{2012MNRAS.422..478R} suggested a correction to bolometric luminosities with the assumption of isotropy. Using a randomly selected quasar sample from \cite{2011Shang} and a preferential view angle of 31\degree\ from \cite{1989Barthel}, \cite{2010Nemmen} suggested bolometric corrections factor $f$ as 0.75. This correction will increase the bolometric luminosity value. We add this correction as an option in the final code.

\subsection{Bolometric Correction and Multi-band Linear Regression}
\label{Sec:BC}

Bolometric correction (BC) is a widely used method to calculate bolometric luminosities for large quasar samples. This traditional BC is given as
\begin{equation}
    \mathrm{BC}_\nu=\frac{L_{\rm{bol}}}{\nu L_\nu}.
\end{equation}
This definition assumes a coefficient of unit between monochromatic luminosity and bolometric luminosity in the logarithmic space. Bolometric correction values from different works are usually different. Table~\ref{tab:wave_bc} summarizes correction values from previous studies, as well as our correction values. The selection of the X-ray wavelength boundaries has little influence on the calculation of the bolometric luminosity, because the X-ray luminosity is approximately one order of magnitude lower than the UV/optical luminosity. We can see that the BC values highly rely on the IR wavelength used for the computation, since the IR emission contributes a large portion of the whole quasar SED. 

\begin{figure}
\centering
\includegraphics[width=0.48\textwidth]{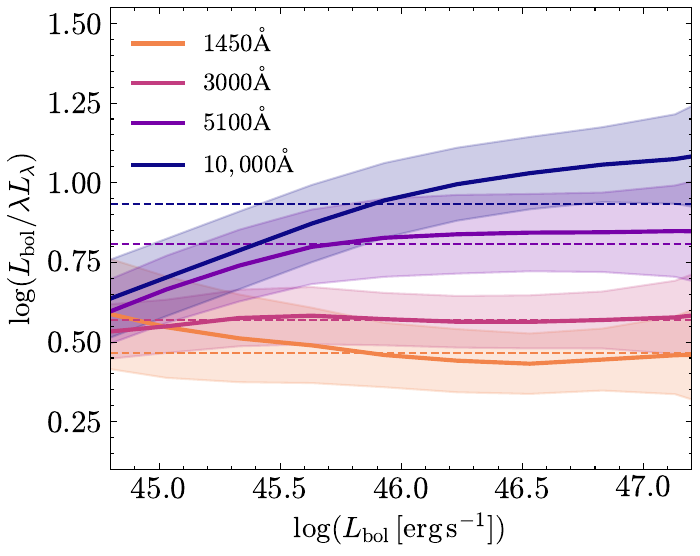}
\caption{Bolometric correction as a function of bolometric luminosity. The dashed horizontal lines represent the median values of the bolometric corrections. }
\label{fig:BC}
\end{figure}

\cite{2010Nemmen} and \cite{2012MNRAS.422..478R} found that the relation between monochromatic luminosity and bolometric luminosity depends on luminosity, and thus used a linear relationship in the logarithmic space. In Figure~\ref{fig:Llambda-Lbol}, we show the relation between bolometric luminosity and monochromatic luminosity at 1450\,\r{A}, 3000\,\r{A}, 5100\,\r{A}, and 10,000\,\r{A} for our sample. We evaluate the bolometric correction as a function of bolometric luminosity in Figure~\ref{fig:BC}. The horizontal dashed lines represent the median BC values for the corresponding wavelengths. The figure clearly displays a luminosity dependence with a large scatter. Specifically, the optical and IR BCs overestimate $L_{\rm{bol}}$ at low luminosities and underestimate $L_{\rm{bol}}$ at high luminosities, while the UV BCs show an opposite trend.

To take advantage of the multi-band data, we use the least square method to fit the relationship between bolometric luminosity and monochromatic luminosity at four wavelengths, i.e., 1450\,\r{A}, 3000\,\r{A}, 5100\,\r{A}, and 10,000\,\r{A}. We optimize the multiple linear regression (MLR) by minimizing $\chi^2$, given by $\chi^2=\sum(y_{i}-y_{i,\mathrm{model}})^2/(\sigma_{y,i} ^2)$, where $y_i$, $y_{i,\mathrm{model}}$, and $\sigma_{y,i}$ are $\log (L_{\rm bol})$, the model $\log (L_{\rm {bol}})$, and the $1\sigma$ uncertainty of $y_i$, respectively.
Table~\ref{tab:bolometric_correction} shows the multiple linear regression results from one monochromatic luminosity. 
Table~\ref{tab:multi_wavelength_correction} shows the multiple linear regression results from two and three monochromatic luminosities. The uncertainties of the slope and intercept are calculated following \cite{Hogg2010} assuming a normal distribution for the computed bolometric luminosities. For each BC, we calculate a fractional residual $f=(L_{\rm{bol, model}}-L_{\rm{bol, true}})/L_{\rm{bol, true}}$ between the predicted bolometric luminosity and the true value. The dispersion of the fractional residual ($\sigma_f$) represents the accuracy of the bolometric luminosity computation. The $\sigma_f$ is estimated as half the difference between the 16th and 84th percentiles of the fractional residual. The $\sigma_f$ values based on one, two, and three monochromatic luminosities are about 24.5\%, 16.7\%, and 13.6\%, respectively.

We also compare monochromatic luminosities with spectra-based luminosities from \cite{2022ApJS..263...42W}. These monochromatic luminosities are roughly the same with a large scatter in the low luminosity range likely due to quasar variability and fiber loss. We perform a test and replace the $ugriz$ photometry with the SDSS spectra, and find a small difference (1\%) in bolometric luminosity with a 7\% scatter. The bolometric corrections from spectral-photometric combined bolometric luminosity have a larger $\sigma_f$ of 26\% than that inferred from broadband SED.

\begin{table*}
\centering
    \caption{Bolometric corrections and comparison with previous results}
    \begin{tabular}{cccccccc}
    \hline \hline
        &This paper  &This paper     &\cite{2006ApJS..166..470R}  &\cite{2012MNRAS.422..478R} & \cite{2013ApJS..206....4K}\\
         \hline 
Integration range& 4\,\micron-10\,keV& 1\,\micron-10\,keV&30\,\micron-10\,keV&1\,\micron-8\,keV&1\,\micron-2\,keV\\
$\mathrm{BC}_{1450}$
&$2.93\pm0.64$&$2.27\pm0.39$& $(2.33\tablenotemark{a})$&$4.2\pm0.1$&\nodata\vspace{0.15cm}\\

$\mathrm{BC}_{3000}$
&$3.72\pm0.68$&$2.86\pm0.63$
&$5.62\pm1.14$&$5.2\pm0.2$&$\mathrm{BC}_{2500}=2.75\pm0.40$ \vspace{0.15cm}\\

$\mathrm{BC}_{5100}$
&$6.43\pm1.84$&$4.98\pm1.73$ 
&$10.3\pm2.1$&$8.1\pm0.4$&$4.33\pm1.29$ ($7.79\pm1.69$\tablenotemark{b})\vspace{0.15cm}\\

$\log (L_{\rm bol})$ range
&44.58-48.07&-
&45.06-47.43&45.13-47.30&45.06-47.43 \\

\hline
    \end{tabular}
    \label{tab:wave_bc}

\tablenotetext{a} {value for the mean SED of all 
quasars in \cite{2006ApJS..166..470R} integrating from 1\,$\mu$m to 8\,keV as reported by \cite{2012MNRAS.422..478R}.}
\tablenotetext{b} {value with integrating from 30\,\micron\ to 10\,keV as reported by \citet{2013ApJS..206....4K}.  }
\end{table*}

\begin{table*}
\centering
\caption{Bolometric corrections from one monochromatic luminosity.}
\label{tab:bolometric_correction}
\begin{tabular}{lccccc}
\hline \hline
Wavelength & $ L_{\rm bol} = \rm{BC_{\lambda}}\, \lambda L_{\lambda}$ & $f$& $\sigma_{\rm f}$ \\
\hline
1450\,\r{A} & $L_{\rm bol} = (2.93 \pm 0.64)\, \lambda L_{\lambda}$ & 0.000& 0.203\\
3000\,\r{A} & $L_{\rm bol} = (3.72 \pm 0.68)\, \lambda L_{\lambda}$ & 0.000& 0.180\\
5100\,\r{A}& $L_{\rm bol} = (6.43 \pm 1.84)\, \lambda L_{\lambda}$ & 0.000& 0.290\\
10,000\,\r{A} &$L_{\rm bol} = (8.57 \pm 2.91)\, \lambda L_{\lambda}$ & 0.000& 0.359\\

\hline
Wavelength & $\log (L_{\rm bol}) = A + B \log (\lambda L_{\lambda})$ & $f$& $\sigma_{\rm f}$ \\
\hline
1450\,\r{A} & $\log (L_{\rm bol}) = (6.020 \pm 0.001) + (0.879 \pm 0.001) \log (\lambda L_{\lambda})$ & 0.068& 0.197\\
3000\,\r{A} & $\log (L_{\rm bol}) = (2.118 \pm 0.001) + (0.966 \pm 0.01) \log (\lambda L_{\lambda})$ & $-0.002$& 0.176\\
5100\,\r{A} & $\log (L_{\rm bol}) = (0.633 \pm 0.001) + (1.004 \pm 0.001) \log (\lambda L_{\lambda})$ & $-0.028$& 0.282\\
10,000\,\r{A} & $\log (L_{\rm bol}) = (-4.101 \pm 0.001) + (1.112 \pm 0.001) \log (\lambda L_{\lambda})$ & $-0.018$& 0.323\\
\hline
\end{tabular}
\tablecomments{Uncertainties for all fits in this table are $1\sigma$ confidence level.}
\end{table*}

\begin{table*}
\centering
\caption{Bolometric corrections from two and three monochromatic luminosities.}
\label{tab:multi_wavelength_correction}
\begin{tabular}{lcccccc}
\hline \hline
Two-Wavelength & \multicolumn{4}{c}{log($L_{\rm bol}$) = $A$ + $B_1$ log($\lambda_1 L_{1}$) + $B_2$ log($\lambda_2 L_{2}$)} & $f$& $\sigma_{\rm f}$\\
\hline

1450\,\r{A}, 3000\,\r{A}   & $A = 4.139 \pm 0.001$ & $B_1 = 0.238 \pm 0.001$ & $B_2 = 0.684 \pm 0.001$ & & 0.020&  $0.145$ \\
1450\,\r{A}, 5100\,\r{A}   & $A = 3.018 \pm 0.001$ & $B_1 = 0.441 \pm 0.001$ & $B_2 = 0.508 \pm 0.001$ &&  0.014& $0.140$ \\
1450\,\r{A}, 10,000\,\r{A}   & $A = 0.571 \pm 0.001$ & $B_1 = 0.489 \pm 0.001$ & $B_2 = 0.515 \pm 0.001$ & &0.019&  $0.136$ \\
3000\,\r{A}, 5100\,\r{A}   & $A = 3.199\pm 0.001$ & $B_1 = 0.804 \pm 0.001$ & $B_2 = 0.140\pm 0.001$ && $0.004$&  $0.167$ \\
3000\,\r{A}, 10,000\,\r{A}   & $A = 1.902 \pm 0.001$ & $B_1 = 0.766\pm 0.001$ & $B_2 = 0.208 \pm 0.001$ & &$0.004$& $0.162$ \\
5100\,\r{A}, 10,000\,\r{A}   & $A = 0.727\pm 0.001$ & $B_1 = 0.757 \pm 0.001$ & $B_2 = 0.246 \pm 0.001$ & & $-0.013$&  $0.254$ \\

\hline
Three-Wavelength & \multicolumn{4}{c}{log($L_{\rm bol}$) = $A$ + $B_1$ log($\lambda_1 L_{1}$) + $B_2$ log($\lambda_2 L_{2}$)+ $B_3$ log($\lambda_3 L_{3}$)}&$f$ & $\sigma_{\rm f}$\\
\hline
1450\,\r{A}, 3000\,\r{A}, 5100\,\r{A}   & $A = 3.159\pm 0.001$ & $B_{1} = 0.342 \pm 0.001$ & $B_{2} = 0.254 \pm 0.001$ & $B_{3} = 0.350 \pm 0.001$ & 0.013& $0.132$ \\
1450\,\r{A}, 3000\,\r{A}, 10,000\,\r{A}   & $A = 1.288\pm 0.001$ & $B_{1} = 0.351 \pm 0.001$ & $B_{2} = 0.276\pm 0.001$ & $B_{3} = 0.360 \pm 0.001$ & 0.017& $0.121$ \\
1450\,\r{A}, 5100\,\r{A}, 10,000\,\r{A}   & $A = 1.215 \pm 0.001$ & $B_{1} = 0.451 \pm 0.001$ & $B_{2} = 0.211 \pm 0.001$ & $B_{3} = 0.328 \pm 0.001$ & 0.016& $0.128$ \\
3000\,\r{A}, 5100\,\r{A}, 10,000\,\r{A}   & $A = 1.749\pm 0.001$ & $B_{1} = 0.807 \pm 0.001$ & $B_{2} = -0.092 \pm 0.001$ & $B_{3} = 0.262 \pm 0.001$ & $0.006$& $0.162$ \\
\hline
\end{tabular}
\tablecomments{Uncertainties for all fits in this table are $1\sigma$ confidence level.}
\end{table*}

\section{Results}
\label{Sec:SOM}

Machine learning (ML) emerges as a transformative tool for multi-dimensional datasets. Combining the strength of photometric surveys with the advanced machine learning tool may offer a pathway to overcome the challenge of calculating robust bolometric luminosities for large numbers of quasars. By learning directly from the data, ML techniques can efficiently process multi-dimensional datasets and uncover complex patterns that are otherwise difficult to model with traditional approaches \citep{2010Ball,2016Acquaviva,2019Baron}. The Self-Organizing Map \citep[SOM;][]{KOHONEN82} is an unsupervised ML algorithm that projects high-dimensional data onto a 2D map that nevertheless preserves the topology. SOM has been widely applied in astrophysics. For instance, \cite{2019Hemmati} and \cite{2024LaTorre} applied SOM to improve photometric redshift estimation for large-scale galaxy surveys. \cite{2024LaTorre_Pacucci} used SOM to constrain the properties of slowly accreting black holes. These studies underscore the flexibility of SOM in addressing complex, high-dimensional challenges in astrophysical researches.

SOM maps a high-dimensional parameter space onto a 2D lattice, where each neuron is associated with a weight vector corresponding to the high-dimensional input space. Here, we use the Python library SomPY\footnote{\url{https://github.com/sevamoo/SOMPY}} \citep{moosavi2014sompy} to construct and train our SOM model. In SomPY, training data are normalized to the unit variance with a zero mean. The training of a SOM begins with initializing the weight vectors using the principal component analysis \citep[PCA;][]{Chatfield1980}, followed by the iterative adjustment of weights. For each data point in the training set, the Best Matching Unit (BMU), the neuron whose weight vector is closest to the input, is identified using a Euclidean distance metric. The BMU's weight vector is updated to move closer to the input data, and a neighborhood function ensures that adjacent neurons in the lattice are similarly adjusted. This iterative process continues until a convergence occurs when the weights stabilize. As an unsupervised method, SOM does not require labeled input data, making it particularly valuable for exploratory data analysis and visualization. We refer to the neuron as a `cell' in a 2D map in the following context.

\subsection{Generating SOM}

\begin{figure}
\centering
\includegraphics[width=0.5\textwidth]{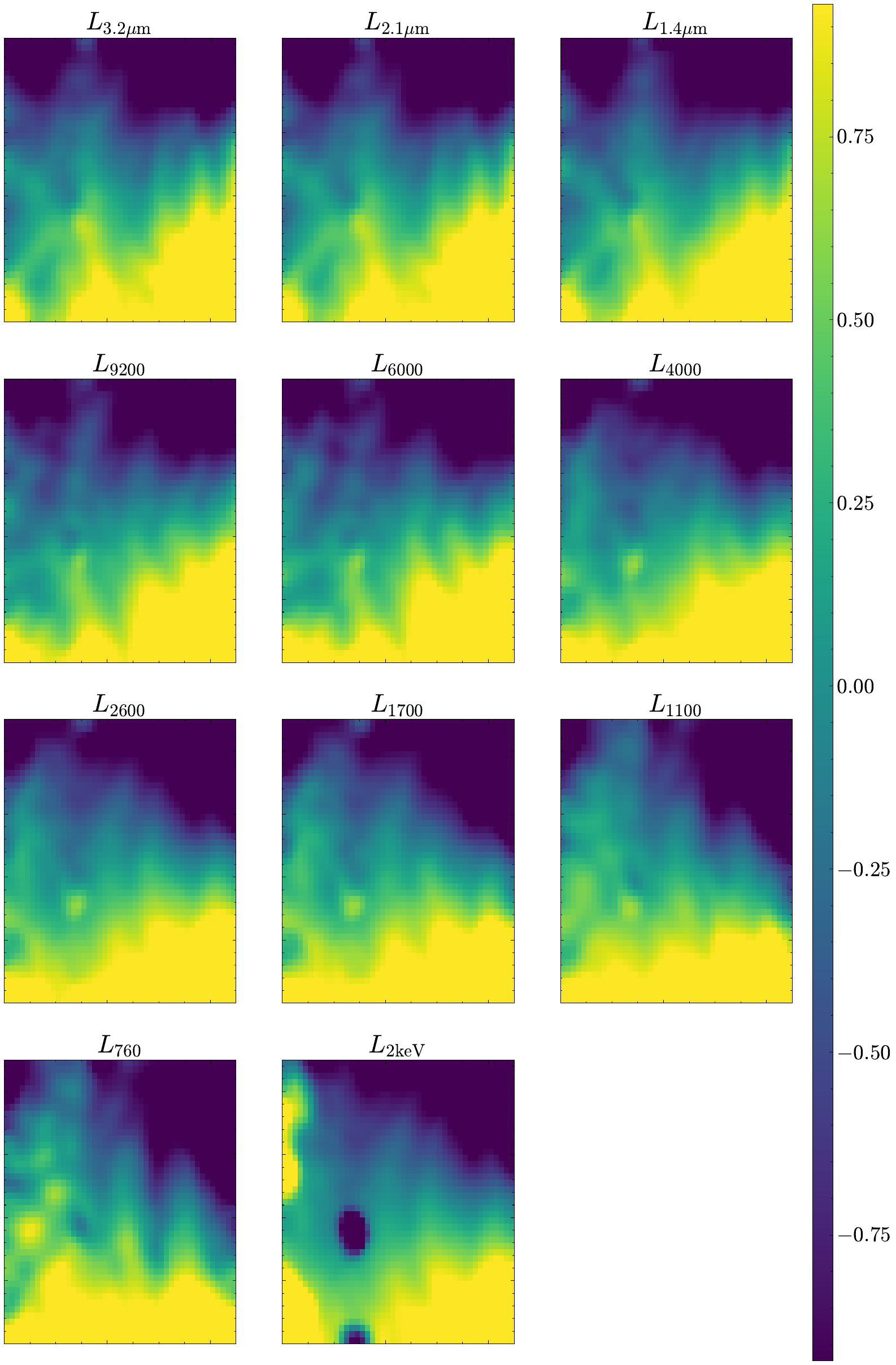}
\caption{Distribution of 11 normalized monochromatic luminosities on the 45$\times$45 trained SOM.}
\label{fig:feature}
\end{figure}

In this work, the input parameter space has 11 dimensions representing the rest-frame quasar SEDs (i.e., 11 monochromatic luminosities). We select monochromatic luminosities in the range of $3.2\,\rm{\micron}$ to 760\,\r{A}, along with the 2 keV luminosity. Our photometric data are available for most quasars in this wavelength range. To evaluate the precision and adaptability of this method, we randomly split our sample into a training set (80\%) and a test set (20\%). 

The SOM performance depends on the map size and geometry \citep[see][]{DAVIDZON19}. The map size and geometry should balance a good sampling of the data and a high resolution. A sample rate is defined as the fraction of the cell with more than 20 samples. The quantization error is the mean of the Euclidean distances of all training data to their BMUs. A smaller quantization error indicates a high resolution. We choose a 45$\times$45 SOM with a sampling rate of 90\% and a quantization error of 0.11. Figure~\ref{fig:feature} shows the distribution of 11 normalized monochromatic luminosities on the 45$\times$45 map. We also verify that the 11 monochromatic luminosity distributions in the input dataset and across the SOM cells are consistent, ensuring that the trained SOM represents the input data.

\subsection{Bolometric Luminosity Projection}

\begin{figure}
\centering
\includegraphics[width=0.5\textwidth]{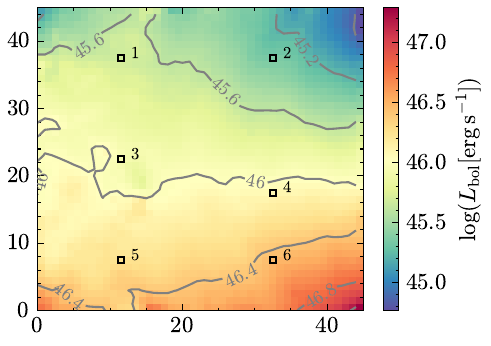}
\includegraphics[width=0.5\textwidth]{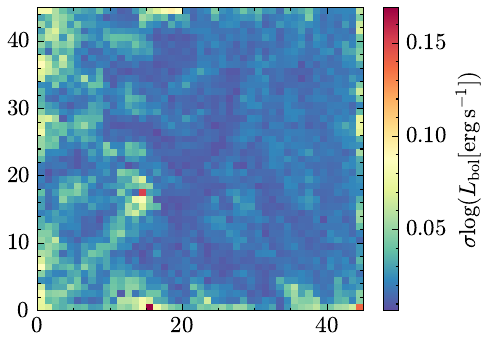}

\caption{Upper panel: SOM with cells color-coded by the median values of $\log (L_{\rm {bol}})$. The gray contours represent the levels of the $\log (L_{\rm {bol}})$ values. The open black squares represent six cells: Cell 1 = [33,\,38], Cell 2 = [12,\,38], Cell 3 = [12,\,23], Cell 4 = [33,\,18], Cell 5 = [12,\,8], Cell 6 = [33,\,8]. Lower panel: SOM with cells color-coded with the $1\sigma$ level of $\log (L_{\rm {bol}})$.}
\label{fig:Lbol map}
\end{figure}

\begin{figure}
\includegraphics[scale=1]{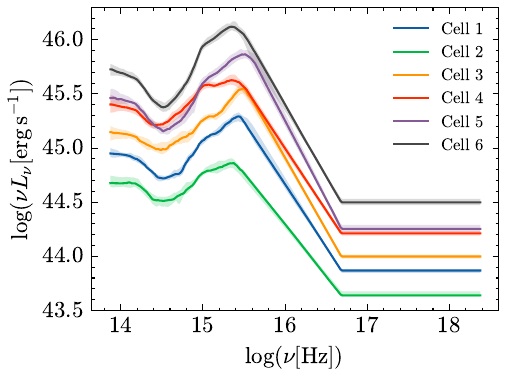}
\caption{Six median quasar SEDs (solid curves) with $1\sigma$ dispersions (shaded area) from six cells, i.e., the black open circles in Figure~\ref{fig:Lbol map}.}
\label{fig:SED map}
\end{figure}

Since bolometric luminosity depends on the SED, objects in the same cell share similar SEDs and bolometric luminosities. We compute the median value and $1\sigma$ deviation (i.e., half the difference between the 84th and 16th percentiles) of $\log (L_{\rm {bol}})$ for each cell. Figure~\ref{fig:Lbol map} shows the trained SOM color-coded by the median and $1\sigma$ deviation of $\log (L_{\rm {bol}})$. The $\log (L_{\rm {bol}})$ values smoothly increase nearly vertically from top to bottom, and the $\sigma( \log L_{\rm{bol}})$ values are globally small, indicating that this SOM model is a good classifier for $\log (L_{\rm {bol}})$. To demonstrate that this SOM describes the SED diversity, we select six SEDs from six cells representing different luminosities in Figure~\ref{fig:SED map}. The shaded region is the $1\sigma$ deviation of the SEDs, suggesting that the SEDs in each cell are well constrained, with small deviations around the median.

\begin{figure}
\centering
\includegraphics[width=0.5\textwidth]{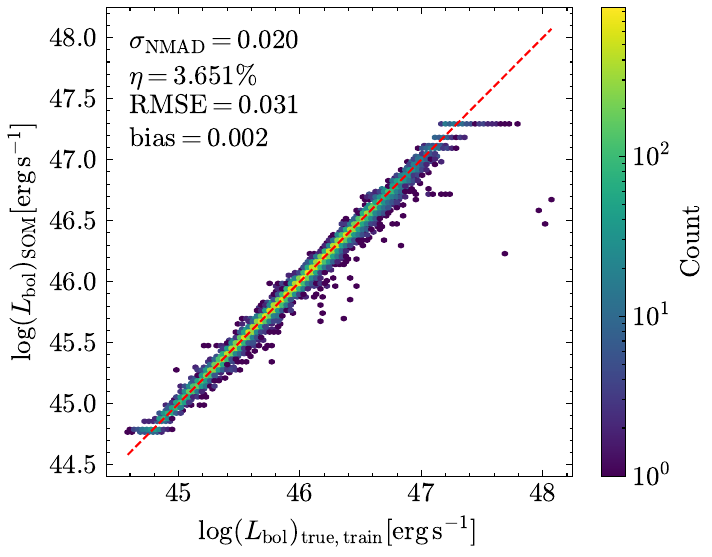}
\includegraphics[width=0.5\textwidth]{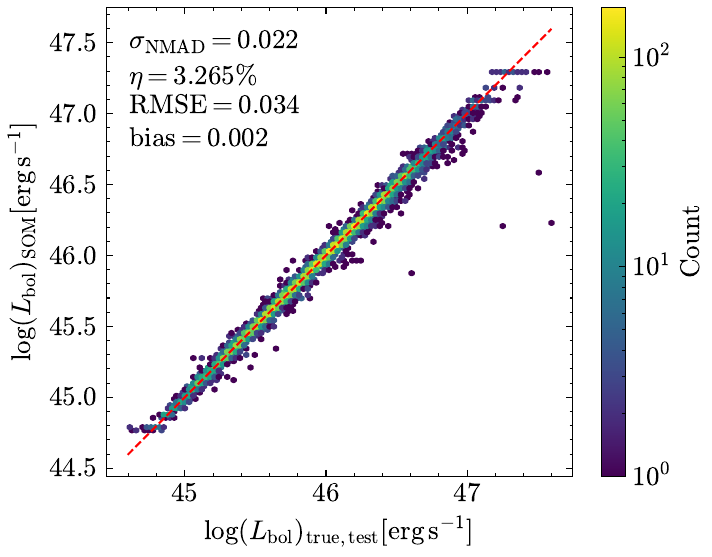}

\caption{Comparison between the true $\log (L_{\rm {bol}})$ and SOM-derived $\log (L_{\rm {bol}})$ values in the training and test sets. The red line is the one-to-one line.}
\label{fig:L-Lsom}
\end{figure}

Figure~\ref{fig:L-Lsom} compares the true values with the SOM-derived values for $\log (L_{\rm {bol}})$ in the training and test samples. We also compute four statistical parameters, including the normalized median absolute deviation ($\sigma_{\rm{NMAD}}$), outlier fraction $\eta$, root mean squared error (RMSE), and bias. The outlier fraction $\eta$ is the fraction with $|\log (L_{\rm{bol, true}}) -\log (L_{\rm{bol, som}})| \geq 2\sigma$, where $\sigma$ is the standard deviation of the $\log (L_{\rm{bol, true}}) -\log (L_{\rm{bol, som}})$ values. With the SOM model, the $\log (L_{\rm {bol}})$ estimation is accurate in the training and test samples, with $\sigma_{\rm{NMAD}}\approx 0.02$ and RMSE$\approx 0.03$. We also use the target selection flags from SDSS DR16Q and separate WISE-selected quasars or X-ray-selected quasars as a test sample. The remaining quasars stay in the training sample. Overall, $\sigma_{\rm{NMAD}}$ and RMSE are largely unaffected by sample selection.
For each quasar, we use the systematic uncertainty, i.e., the dispersion of $\log (L_{\rm {bol}})$ within the best-fit cell.

We project other fundamental AGN parameters (black hole mass and Eddington ratio) in the SED-based SOM model. They have worse correlations with the SED compared to the bolometric luminosity. It is likely due to large uncertainties in the current estimate of single-epoch black hole masses, or an intrinsic weak correlation with SED \citep{2020A&A...636A..73D}.

\subsection{Comparing with the Bolometric Correction Method}

\begin{figure}

\centering
\includegraphics[width=0.5\textwidth]{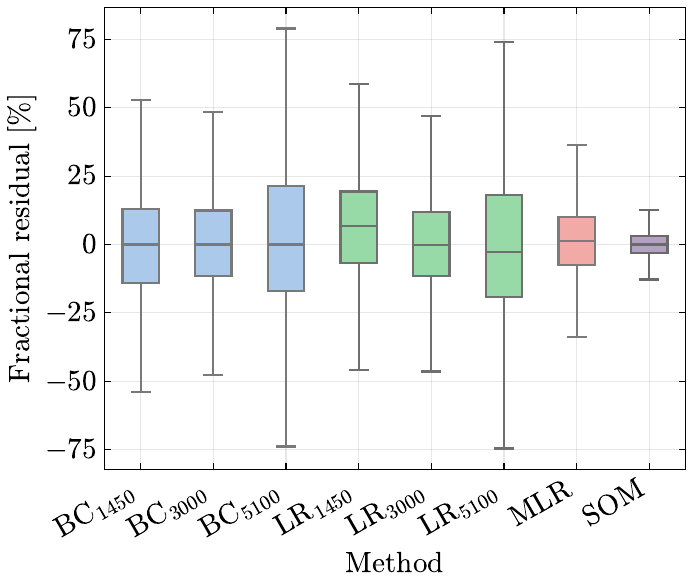}
\caption{Fractional residuals from different methods, including BCs at 1450\,\r{A}, 3000\,\r{A}, and 5100\,\r{A}, linear regression with monochromatic luminosities at 1450\,\r{A}, 3000\,\r{A}, and 5100\,\r{A}, multiple linear regression with monochromatic luminosities at 1450\,\r{A}, 3000\,\r{A}, and 5100\,\r{A}, and the SOM method. For clarity, outliers are not shown.}
\label{fig:f}
\end{figure}

We calculate the median and dispersion of fractional residuals between the SOM-derived bolometric luminosity and the true value. For comparison, we also use the BCs and linear relations with monochromatic luminosities to predict bolometric luminosities. The distributions of the fractional residuals from different methods are shown in Figure~\ref{fig:f}, where the SOM method result is from all samples. The line in the box represents the median, while the top and bottom edges correspond to the 75th and 25th percentiles. The whiskers extend to the minimum and maximum values. The median values of the fractional residuals are near zero in all methods. The SOM method with 11 monochromatic luminosities has the smallest dispersion relative to other methods, with $\sigma_f$ of 4.9\%. For the method with BC, $\sigma_f$ is about 20.3\%, 18.0\%, and 29.0\% at 1450\,\r{A}, 3000\,\r{A}, and 5100\,\r{A}, respectively. For the method of the linear fitting with monochromatic luminosity, $\sigma_f$ is about 19.7\%, 17.6\%, and 28.2\% at 1450\,\r{A}, 3000\,\r{A}, and 5100\,\r{A}, respectively. In short, the SOM method with 11 monochromatic luminosities has largely reduced the dispersion of the fractional residual. 

\subsection{Handling Limited Data}
\label{subsec:KNN}

We also use the SOM model to compute the bolometric luminosities of quasars with limited data. As we mentioned earlier, large-area surveys such as SDSS, UKIDSS, and WISE covered most of the currently known quasars, so these quasars have photometric data from UV to mid-IR. Despite the fact, many quasars have fewer photometric data points available. Therefore, we explore how to use SOM to deal with limited data.
\cite{2024LaTorre} proposed a method to recover specific missing information by randomly drawing 5000 new values of the missing feature from the distribution of the rest of the training sample, under the assumption that the missing value obeys the same distribution as the training sample. Given the new set of input features, they created a 2D likelihood surface of possible cells and extracted the probability distribution function of the interested property. Since the features in SEDs are highly correlated, randomly drawing new values may produce non-physical SEDs. Hence, we use \texttt{KNNImputer} \citep{Troyanskaya}, an imputation for completing missing values using k-Nearest Neighbors, to recover missing data. Each sample's missing values are imputed using the mean value of the nearest $n$ neighbors found in the training set. We set $n\_neighbors=100$ in \texttt{KNNImputer}. After recovering limited SEDs, we use the fully-trained SOM model to compute bolometric luminosities. 

\begin{figure}

\centering
\includegraphics[width=0.48\textwidth]{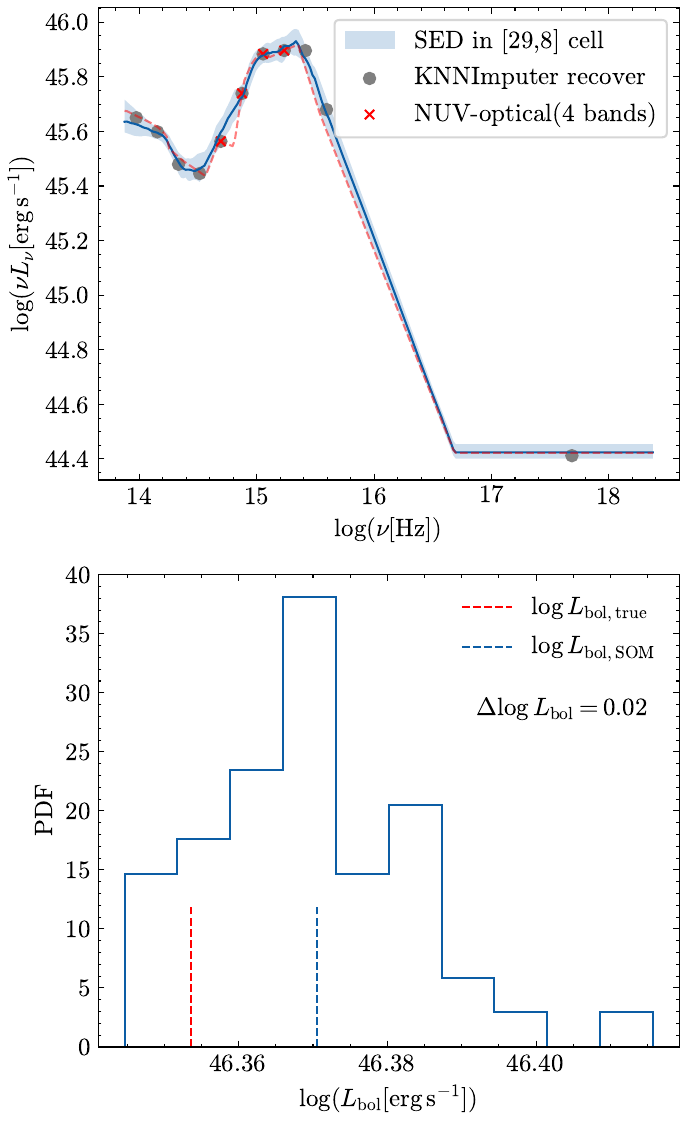}

\caption{An example for \texttt{KNNImputer} to recover limited SEDs. The upper panel shows input and recovered SEDs. The input data only contains 4 band luminosities from NUV to optical (red crosses). The gray dots denote the recovered SED by \texttt{KNNImputer}, and the red dashed curve represents the true SED of this quasar. The blue curve and shaded area are the median and $1\sigma$ SED from cell [29,\,8] in SOM, where the recovered SED belongs. The lower panel shows the $\log (L_{\rm {bol}})$ distribution in cell [29,\,8]. The blue and red dashed line represents the median $\log (L_{\rm {bol}})$ value in cell [29,\,8] and the true $\log (L_{\rm {bol}})$ value of this quasar. The difference $\Delta \log L_{\rm{bol}}=\log (L_{\rm{bol,SOM}})-\log (L_{\rm{bol,true}})$ is $0.02$.}
\label{fig:reparie-SED}
\end{figure}

\begin{figure}

\centering
\includegraphics[width=0.5\textwidth]{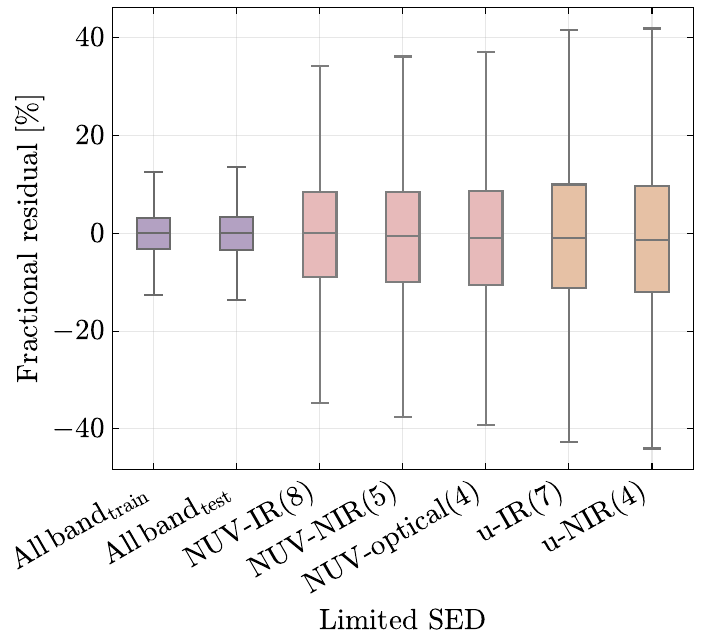}
\caption{Fractional residuals for the limited SEDs with the SOM method. For clarity, outliers are not shown.}
\label{fig:f-limited}
\end{figure}

We use the test sample as an example to demonstrate the capability of \texttt{KNNImputer}. We select five typical limited SEDs that mimic the absence of real observations. These consecutive SEDs are divided into two categories, where SEDs start from 1700\,\r{A} and 2600\,\r{A}, respectively. Table \ref{tab:uncertainty} lists the details, including the number of photometric data points. Figure~\ref{fig:reparie-SED} shows an example of applying \texttt{KNNImputer} to recover limited SEDs. The fractional residuals after recovering limited SEDs are shown in Figure~\ref{fig:f-limited}. Compared with fractional residuals using all bands above, a limited SED has a higher fractional residual, because less information introduces a larger uncertainty. The results also indicate that the UV part of the SED is important to reduce estimator uncertainties. We also test limited SEDs without recovery and find that the closest BMU ignores the missing features. The $1\sigma $ confidence level of the fractional residual increases about 5\% for the five limited SEDs, suggesting that recovery is necessary.

\begin{table}
\centering
\caption{Effect of bolometric luminosity recovery for different missing bands}
\begin{tabular}{lll}
\hline
\hline
Limited SED & f& $\sigma_f$ \\
\hline
All bands\,(all sample)   &   0.000  & 0.049  \\
All bands\,(training set)   &   0.000  & 0.048  \\
All bands\,(test set)   &   0.001 & 0.052  \\
NUV-IR\,(8 bands)   &   0.001 & 0.129  \\
NUV-NIR\,(5 bands)   &   $-0.005$ & 0.137  \\
NUV-optical\,(4 bands)   &$-0.008$ & 0.143  \\
u-IR\,(7 bands)    &   $-0.010$ & 0.159  \\
u-NIR\,(5 bands)  &   $-0.014$ & 0.160  \\
\hline
\end{tabular}
\label{tab:uncertainty}
\end{table}

\subsection{Application to SDSS DR16Q}
\label{sub:application_to_sdss_dr16q}
We apply our method to the remaining quasars in the SDSS DR16Q sample and calculate their $\log (L_{\rm {bol}})$. We focus on typical type 1 luminous quasars with $\log(L_{2500}/\rm{erg\,s^{-1}})>43.7$ at $z<5$ and obtain 633,373 sources. We crossmatch with the existing surveys to obtain photometric data across UV to near-IR following the same method in section~\ref{Sec:Data}. For W1 and W2 in the mid-IR, we use unWISE \citep{Schlafly2019} to increase the detection rate. For the IGM absorption correction, we use the average IGM transmission curves for FUV, NUV, $u$, $g$, and $r$ bands for redshift from 0 to 5. Then we constructed restframe SEDs with photometric points with S/Ns greater than 3 and interpolated at 10 features corresponding to wavelengths from 760\,\r{A} to 3.2\,\micron. We also obtain the X-ray emission of the quasars using the same method earlier. 

If a quasar has any missing features, we recover them using the full-coverage samples with \texttt{KNNImputer} introduced in Section~\ref{subsec:KNN}, i.e., the average of the 100 most similar SEDs for recovery. With the recovered SED, the SOM can find the most similar subsample of the target. As we discussed in Section~\ref{Sec:SOM} that less information induces a larger uncertainty, here we use the modeling systematic uncertainty in the test sample with limited data from Section~\ref{subsec:KNN} as the uncertainty of predicted $\log (L_{\rm {bol}})$. We publicly release bolometric luminosities for these recalculated DR16Q samples in Table~\ref{tab:quasar_catalog} in Appendix~\ref{appendixA}.

\begin{figure}
\centering
\includegraphics[width=0.48\textwidth]{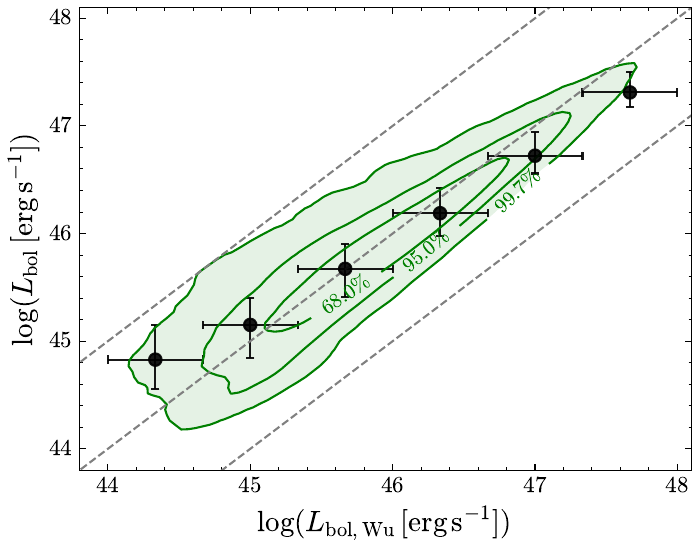}
\caption{Comparison between $\log (L_{\rm {bol}})$ from this work and $\log (L_{\rm{bol,\,Wu}})$ from \cite{2022ApJS..263...42W} for recalculated DR16Q sample. The contour level is labeled with a closed percentile. The gray dashed lines, from top to bottom, represent a line one dex higher, the one-to-one line, and a line one dex lower.}
\label{fig:bol-z}
\end{figure}

For quasars from 79,163 full-coverage samples and 633,373 suitable quasar samples, we compare their bolometric luminosity distribution with those from \cite{2022ApJS..263...42W}, who mainly used $\rm{BC}=5.15$ at 3000\,\r{A} in Figure~\ref{fig:bol-z}. Additionally, we bin the sample into six equally bins of $\log (L_{\rm{bol,\,Wu}})$, and display the corresponding median values for each bin in Figure~\ref{fig:bol-z}. Their $x$-axis error indicates the bin size, and their $y$-axis error indicates the 16th to 84th percentiles of the isotropic bolometric luminosity based on the SOM method. In summary, the median and $1\sigma$ uncertainty (i.e., half the difference between the 84th and 16th percentiles) values of $(L_{\rm{bol,\,Wu}} - L_{\rm{bol}})/L_{\rm{bol}}$ are 14.8\% and 43.6\%, respectively.

\subsection{Code Description}

We make our Python code publicly available\footnote{\url{https://github.com/ChenJiemi/QSOLbol}}, including the fully-trained SOM model, training and test data sets, luminosity maps, and modeling systematic uncertainties. To compute the bolometric luminosity of a quasar, users will provide its redshift and observed-frame SED. The usage is explained by an example in the code package. The parameters of the main function \texttt{QSOLbol} are listed in Table~\ref{tab: param}. Users can modify the parameter $f\_isotropy$ to control the viewing angle correction from \cite{2012MNRAS.422..478R}. The wavelength range of bolometric luminosity can be generated according to users' requirements, and the range of the selection should be in the range of 4\,\micron-10 keV. 

Since our samples are small in number at the faint and bright luminosity ends, we add a scaling parameter to scale up or scale down the SEDs of the faint ($\log (L_{\rm {bol}}/\rm{erg\,s^{-1}})<45.13$) or bright sources ($\log (L_{\rm {bol}}/\rm{erg\,s^{-1}})>46.74$). We use the SOM to calculate the bolometric luminosity iteratively. This approach allows for a more accurate calculation of the bolometric luminosity for the bright and faint end sources. We add this scaling parameter as an option in the code. The function returns $\log (L_{\rm {bol}})$, its corresponding uncertainty and scaling factor following the procedures in Section~\ref{sub:application_to_sdss_dr16q}.

\begin{table*}
\centering
\caption{Parameters in \texttt{QSOLbol}}
\begin{tabular}{lll}
\hline
\hline
Parameters& Type& Description \\
\hline
wave  &  array-like &  Observed-frame SED wavelength in Angstrom.\\
mags  &   array-like & Observed-frame SED in AB magnitude.  \\
mags\_err  &   array-like & Observed-frame SED error in AB magnitude. \\
z  &   1D array-like  & Redshift. \\
f\_isotropy  &bool& Whether to use 0.75 correction for viewing angle. The default is False.\\
wave\_range  &tuple  & The integrated range of bolometric luminosity in Hz. The default is from 4\,\micron\,to 2 keV. \\
scale &bool & Whether to scale the SED when the source is at faint and bright luminosity ends. \\
\hline
\end{tabular}
\label{tab: param}
\end{table*}

\section{Summary}
\label{Sec:Summary}

We have developed a new method to estimate quasar isotropic bolometric luminosities. Our approach is based on a sample of 79,163 quasars in the redshift range of 0.5 to 2 from SDSS DR16Q. These quasars have multi-wavelength data spanning the wavelength range from the mid-IR to the UV (Table~\ref{tab:quasar_catalog}). After correcting the IGM absorption and host galaxy contamination, we constructed SEDs and computed their bolometric luminosities. The main results are summarized below.

\begin{enumerate}
\item  We constructed a median SED of these quasars, with a restframe coverage from 4\,\micron\,to 2\,keV. The normalized SED clearly shows the diversity of quasar SEDs (see Figures~\ref{fig:SED} and \ref{fig:SED-diversity}; Section~\ref{Sec:SED}).

\item  We computed the bolometric correction and multi-band linear regression for the quasar sample. These relations show a large scatter with $\sigma_f$ about 13\% to 36\% (see Figure~\ref{fig:Llambda-Lbol} and Tables~\ref{tab:wave_bc}, \ref{tab:bolometric_correction} and \ref{tab:multi_wavelength_correction}; Section~\ref{Sec:SED}).

\item We used an unsupervised neural network SOM to describe the SED diversity and used a fully-train SOM model to compute quasar bolometric luminosities. With full coverage SED features, i.e., 11 restframe monochromatic luminosities from ~3.2\,\micron\, to 2\,keV, our bolometric luminosity computation is robust and efficient, with $\sigma_f$ about 4.9\% (see Figures~\ref{fig:Lbol map},~\ref{fig:L-Lsom}, and \ref{fig:f}; Section~\ref{Sec:SOM}).

\item For quasars with limited SEDs, we used \texttt{KNNImputer} to recover the missing features. Briefly, for a limited SED with given features, the most similar SED is found in the training set samples, and the missing features are replaced with the average of the 100 most similar SEDs. Hence, a quasar with a limited SED can also be applied in the SOM model to compute its bolometric luminosity (see Figures~\ref{fig:reparie-SED} and \ref{fig:f-limited}; Section~\ref{Sec:SOM}).

\item We applied this SOM method to the SDSS DR16Q quasars and calculated their bolometric luminosities (see Figure~\ref{fig:bol-z}; Section~\ref{Sec:SOM}).

\end{enumerate}

We have demonstrated that combining the strength of photometric surveys with the advanced machine learning tool SOM offers a pathway to overcome the challenge of calculating robust bolometric luminosities for large numbers of quasars. The accuracy of individual source bolometric luminosity can be further improved by reliable measurements of AGN host galaxies and simultaneous observations. Future deep ultraviolet facilities such as the China Space Station Telescope would prompt the quasar EUV SED study for bolometric luminosity computation and also the understanding of the quasar central engine. In future work, we will further explore the effect of the unseen EUV part of the SED, host galaxy decomposition, and non-simultaneous effects, and characterize these systematic uncertainties in the bolometric luminosity.

\begin{acknowledgments}
We thank Bernie Shiao and Yuhan Wen for the helpful discussion in GALEX cross-matching. We thank Zhiwei Pan and Gaocheng Yin for their suggestions. We acknowledge support from the National Key R\&D Program of China (2022YFF0503401) and the National Science Foundation of China (12225301, 12322303). 
\end{acknowledgments}

\software{AstroML \citep{VanderPlas2012}, Jupyter notebooks \citep{Perez2007}, Matplotlib \citep{Hunter2007}, Numpy \& Scipy \citep{scipy}, SomPY \citep{moosavi2014sompy}, SCIKIT-LEARN \citep{Pedregosa2011}.}

\appendix
\section{Catalog for SDSS DR16Q}
We present the multi-band photometry and bolometric luminosities of the SDSS DR16Q quasars that satisfy $z<5$ and $\log(L_{2500}/\rm{erg\,s^{-1}})>43.7$ in Table~\ref{tab:quasar_catalog}. The uncertainty of each quantity is also included in the catalog. This catalog includes 79,163 full-coverage quasars and 633,373 suitable quasars. SDSS and GALEX magnitudes are in the AB system. UHS, UKIDSS, VHS, and WISE magnitudes are in the Vega system.

\label{appendixA}
\begin{deluxetable}{cll}[h!]
\tablecaption{Multiwavelength Quasar Catalog Format \label{tab:quasar_catalog}}
\tablehead{
\colhead{Column} & \colhead{Name} & \colhead{Description}
}
\startdata
1  & R.A.                  & Right ascension in decimal degrees (J2000) \\
2  & Decl.                 & Declination in decimal degrees (J2000) \\
3  & Redshift            & Spectroscopic redshift from SDSS DR16Q\\
4  & X\_RAY\_MATCHED     & X-ray matching flag\\
4  & Chandra\_FLUX       & X-ray flux from Chandra (rest-frame 2 keV) in erg/s/$\rm{cm}^2$/keV\\
5  & eROSITA\_FLUX       & X-ray flux from eROSITA (rest-frame 2 keV) in erg/s/$\rm{cm}^2$/keV\\
6  & XMM\_FLUX           & X-ray flux from XMM-Newton (rest-frame 2 keV) in erg/s/$\rm{cm}^2$/keV\\
7  & GALEX\_FUV\_MAG     & GALEX FUV magnitude \\
8  & GALEX\_NUV\_MAG     & GALEX NUV magnitude \\
9  & SDSS\_u\_MAG        & SDSS $u$-band PSF magnitude \\
10 & SDSS\_g\_MAG        & SDSS $g$-band PSF magnitude \\
11 & SDSS\_r\_MAG        & SDSS $r$-band PSF magnitude \\
12 & SDSS\_i\_MAG        & SDSS $i$-band PSF magnitude \\
13 & SDSS\_z\_MAG        & SDSS $z$-band PSF magnitude \\
14 & UKIRT\_Y\_MAG      & UHS and UKIDSS Y-band magnitude \\
15 & UKIRT\_J\_MAG      & UHS and UKIDSS J-band magnitude \\
16 & UKIRT\_H\_MAG      & UHS and UKIDSS H-band magnitude \\
17 & UKIRT\_K\_MAG      & UHS and UKIDSS K-band magnitude \\
18 & VHS\_Y\_MAG         & VHS Y-band magnitude \\
19 & VHS\_J\_MAG         & VHS J-band magnitude \\
20 & VHS\_H\_MAG         & VHS H-band magnitude \\
21 & VHS\_K\_MAG         & VHS K-band magnitude \\
22 & WISE\_W1\_MAG       & WISE W1-band magnitude \\
23 & WISE\_W2\_MAG       & WISE W2-band magnitude \\
24 & WISE\_W3\_MAG       & WISE W3-band magnitude \\
25 & WISE\_W4\_MAG       & WISE W4-band magnitude \\
26 & logLbol             & Bolometric luminosity in log for full-coverage samples\\
27 & logLbol\_SOM         & Bolometric luminosity in log from SOM method\\
28 & logLbol\_Wu             & Bolometric luminosity in log from \cite{2022ApJS..263...42W}.\\
\enddata
\tablecomments{This table is available in a machine-readable format in the online journal.}
\end{deluxetable}



\bibliography{reference}

\begin{thebibliography}{}
\expandafter\ifx\csname natexlab\endcsname\relax\def\natexlab#1{#1}\fi
\providecommand{\url}[1]{\href{#1}{#1}}
\providecommand{\dodoi}[1]{doi:~\href{http://doi.org/#1}{\nolinkurl{#1}}}
\providecommand{\doeprint}[1]{\href{http://ascl.net/#1}{\nolinkurl{http://ascl.net/#1}}}
\providecommand{\doarXiv}[1]{\href{https://arxiv.org/abs/#1}{\nolinkurl{https://arxiv.org/abs/#1}}}

\bibitem[{{Acquaviva}(2016)}]{2016Acquaviva}
{Acquaviva}, V. 2016, \mnras, 456, 1618, \dodoi{10.1093/mnras/stv2703}

\bibitem[{{Assef} {et~al.}(2010){Assef}, {Kochanek}, {Brodwin}, {Cool},
  {Forman}, {Gonzalez}, {Hickox}, {Jones}, {Le Floc'h}, {Moustakas}, {Murray},
  \& {Stern}}]{2010Assef}
{Assef}, R.~J., {Kochanek}, C.~S., {Brodwin}, M., {et~al.} 2010, \apj, 713,
  970, \dodoi{10.1088/0004-637X/713/2/970}

\bibitem[{{Auge} {et~al.}(2023){Auge}, {Sanders}, {Treister}, {Urry},
  {Kirkpatrick}, {Cappelluti}, {Ananna}, {Boquien}, {Balokovi{\'c}}, {Civano},
  {Coleman}, {Ghosh}, {Kartaltepe}, {Koss}, {LaMassa}, {Marchesi}, {Peca},
  {Powell}, {Trakhtenbrot}, \& {Turner}}]{2023ApJ...957...19A}
{Auge}, C., {Sanders}, D., {Treister}, E., {et~al.} 2023, \apj, 957, 19,
  \dodoi{10.3847/1538-4357/acf21a}

\bibitem[{{Ball} \& {Brunner}(2010)}]{2010Ball}
{Ball}, N.~M., \& {Brunner}, R.~J. 2010, International Journal of Modern
  Physics D, 19, 1049, \dodoi{10.1142/S0218271810017160}

\bibitem[{{Baron}(2019)}]{2019Baron}
{Baron}, D. 2019, arXiv e-prints, arXiv:1904.07248,
  \dodoi{10.48550/arXiv.1904.07248}

\bibitem[{{Barthel}(1989)}]{1989Barthel}
{Barthel}, P.~D. 1989, \apj, 336, 606, \dodoi{10.1086/167038}

\bibitem[{{Bianchi}(2014)}]{2014Bianchi}
{Bianchi}, L. 2014, \apss, 354, 103, \dodoi{10.1007/s10509-014-1935-6}

\bibitem[{{Bianchi} {et~al.}(2017){Bianchi}, {Shiao}, \&
  {Thilker}}]{2017Bianchi}
{Bianchi}, L., {Shiao}, B., \& {Thilker}, D. 2017, \apjs, 230, 24,
  \dodoi{10.3847/1538-4365/aa7053}

\bibitem[{{Brightman} {et~al.}(2017){Brightman}, {Balokovi{\'c}}, {Ballantyne},
  {Bauer}, {Boorman}, {Buchner}, {Brandt}, {Comastri}, {Del Moro}, {Farrah},
  {Gandhi}, {Harrison}, {Koss}, {Lanz}, {Masini}, {Ricci}, {Stern},
  {Vasudevan}, \& {Walton}}]{2017ApJ...844...10B}
{Brightman}, M., {Balokovi{\'c}}, M., {Ballantyne}, D.~R., {et~al.} 2017, \apj,
  844, 10, \dodoi{10.3847/1538-4357/aa75c9}

\bibitem[{Bruursema {et~al.}(2023)Bruursema, Vrba, Munn, Dorland, Hodapp,
  Varricatt, Kerr, Irwin, \& Lawrence}]{Bruursema2023USNO}
Bruursema, J., Vrba, F., Munn, J., {et~al.} 2023, Bulletin of the AAS, 55

\bibitem[{{Cai} \& {Wang}(2023)}]{2023NatAs...7.1506C}
{Cai}, Z.-Y., \& {Wang}, J.-X. 2023, Nature Astronomy, 7, 1506,
  \dodoi{10.1038/s41550-023-02088-5}

\bibitem[{{Calistro Rivera} {et~al.}(2016){Calistro Rivera}, {Lusso},
  {Hennawi}, \& {Hogg}}]{2016ApJ...833...98C}
{Calistro Rivera}, G., {Lusso}, E., {Hennawi}, J.~F., \& {Hogg}, D.~W. 2016,
  \apj, 833, 98, \dodoi{10.3847/1538-4357/833/1/98}

\bibitem[{{Camarota} \& {Holberg}(2014)}]{2014MNRAS.438.3111C}
{Camarota}, L., \& {Holberg}, J.~B. 2014, \mnras, 438, 3111,
  \dodoi{10.1093/mnras/stt2422}

\bibitem[{Chatfield \& Collins(1980)}]{Chatfield1980}
Chatfield, C., \& Collins, A.~J. 1980, Principal component analysis (Boston,
  MA: Springer US), 57--81, \dodoi{10.1007/978-1-4899-3184-9_4}

\bibitem[{{Cutri} {et~al.}(2021){Cutri}, {Wright}, {Conrow}, {Fowler},
  {Eisenhardt}, {Grillmair}, {Kirkpatrick}, {Masci}, {McCallon}, {Wheelock},
  {Fajardo-Acosta}, {Yan}, {Benford}, {Harbut}, {Jarrett}, {Lake}, {Leisawitz},
  {Ressler}, {Stanford}, {Tsai}, {Liu}, {Helou}, {Mainzer}, {Gettngs},
  {Gonzalez}, {Hoffman}, {Marsh}, {Padgett}, {Skrutskie}, {Beck}, {Papin}, \&
  {Wittman}}]{2014yCat.2328....0C}
{Cutri}, R.~M., {Wright}, E.~L., {Conrow}, T., {et~al.} 2021, {VizieR Online
  Data Catalog: AllWISE Data Release (Cutri+ 2013)}, VizieR On-line Data
  Catalog: II/328. Originally published in: IPAC/Caltech (2013)

\bibitem[{{Davidzon} {et~al.}(2019){Davidzon}, {Laigle}, {Capak}, {Ilbert},
  {Masters}, {Hemmati}, {Apostolakos}, {Coupon}, {de la Torre}, \&
  {Devriendt}}]{DAVIDZON19}
{Davidzon}, I., {Laigle}, C., {Capak}, P.~L., {et~al.} 2019, arXiv e-prints,
  arXiv:1905.13233.
\newblock \doarXiv{1905.13233}

\bibitem[{{Duras} {et~al.}(2020){Duras}, {Bongiorno}, {Ricci}, {Piconcelli},
  {Shankar}, {Lusso}, {Bianchi}, {Fiore}, {Maiolino}, {Marconi}, {Onori},
  {Sani}, {Schneider}, {Vignali}, \& {La Franca}}]{2020A&A...636A..73D}
{Duras}, F., {Bongiorno}, A., {Ricci}, F., {et~al.} 2020, \aap, 636, A73,
  \dodoi{10.1051/0004-6361/201936817}

\bibitem[{{Dye} {et~al.}(2018){Dye}, {Lawrence}, {Read}, {Fan}, {Kerr},
  {Varricatt}, {Furnell}, {Edge}, {Irwin}, {Hambly}, {Lucas}, {Almaini},
  {Chambers}, {Green}, {Hewett}, {Liu}, {McGreer}, {Best}, {Zhang}, {Sutorius},
  {Froebrich}, {Magnier}, {Hasinger}, {Lederer}, {Bold}, \& {Tedds}}]{2018Dye}
{Dye}, S., {Lawrence}, A., {Read}, M.~A., {et~al.} 2018, \mnras, 473, 5113,
  \dodoi{10.1093/mnras/stx2622}

\bibitem[{{Elvis} {et~al.}(1994){Elvis}, {Wilkes}, {McDowell}, {Green},
  {Bechtold}, {Willner}, {Oey}, {Polomski}, \& {Cutri}}]{1994ApJS...95....1E}
{Elvis}, M., {Wilkes}, B.~J., {McDowell}, J.~C., {et~al.} 1994, \apjs, 95, 1,
  \dodoi{10.1086/192093}

\bibitem[{{Evans} {et~al.}(2024){Evans}, {Evans}, {Mart{\'\i}nez-Galarza},
  {Miller}, {Primini}, {Azadi}, {Burke}, {Civano}, {D'Abrusco}, {Fabbiano},
  {Graessle}, {Grier}, {Houck}, {Lauer}, {McCollough}, {Nowak}, {Plummer},
  {Rots}, {Siemiginowska}, \& {Tibbetts}}]{2024ApJS..274...22E}
{Evans}, I.~N., {Evans}, J.~D., {Mart{\'\i}nez-Galarza}, J.~R., {et~al.} 2024,
  \apjs, 274, 22, \dodoi{10.3847/1538-4365/ad6319}

\bibitem[{{Faucher-Gigu{\`e}re}(2020)}]{2020Faucher}
{Faucher-Gigu{\`e}re}, C.-A. 2020, \mnras, 493, 1614,
  \dodoi{10.1093/mnras/staa302}

\bibitem[{{Fitzpatrick}(1999)}]{1999PASP..111...63F}
{Fitzpatrick}, E.~L. 1999, \pasp, 111, 63, \dodoi{10.1086/316293}

\bibitem[{{Fukugita} {et~al.}(1996){Fukugita}, {Ichikawa}, {Gunn}, {Doi},
  {Shimasaku}, \& {Schneider}}]{1996AJ....111.1748F}
{Fukugita}, M., {Ichikawa}, T., {Gunn}, J.~E., {et~al.} 1996, \aj, 111, 1748,
  \dodoi{10.1086/117915}

\bibitem[{{George} {et~al.}(2000){George}, {Turner}, {Yaqoob}, {Netzer},
  {Laor}, {Mushotzky}, {Nandra}, \& {Takahashi}}]{2000ApJ...531...52G}
{George}, I.~M., {Turner}, T.~J., {Yaqoob}, T., {et~al.} 2000, \apj, 531, 52,
  \dodoi{10.1086/308461}

\bibitem[{{Hemmati} {et~al.}(2019){Hemmati}, {Capak}, {Masters}, {Davidzon},
  {Dor{\`e}}, {Kruk}, {Mobasher}, {Rhodes}, {Scolnic}, \&
  {Stern}}]{2019Hemmati}
{Hemmati}, S., {Capak}, P., {Masters}, D., {et~al.} 2019, \apj, 877, 117,
  \dodoi{10.3847/1538-4357/ab1be5}

\bibitem[{{HI4PI Collaboration} {et~al.}(2016){HI4PI Collaboration}, {Ben
  Bekhti}, {Fl{\"o}er}, {Keller}, {Kerp}, {Lenz}, {Winkel}, {Bailin},
  {Calabretta}, {Dedes}, {Ford}, {Gibson}, {Haud}, {Janowiecki}, {Kalberla},
  {Lockman}, {McClure-Griffiths}, {Murphy}, {Nakanishi}, {Pisano}, \&
  {Staveley-Smith}}]{2016A&A...594A.116H}
{HI4PI Collaboration}, {Ben Bekhti}, N., {Fl{\"o}er}, L., {et~al.} 2016, \aap,
  594, A116, \dodoi{10.1051/0004-6361/201629178}

\bibitem[{{Hogg} {et~al.}(2010){Hogg}, {Bovy}, \& {Lang}}]{Hogg2010}
{Hogg}, D.~W., {Bovy}, J., \& {Lang}, D. 2010, arXiv e-prints, arXiv:1008.4686,
  \dodoi{10.48550/arXiv.1008.4686}

\bibitem[{{Hopkins} {et~al.}(2007){Hopkins}, {Richards}, \&
  {Hernquist}}]{2007ApJ...654..731H}
{Hopkins}, P.~F., {Richards}, G.~T., \& {Hernquist}, L. 2007, \apj, 654, 731,
  \dodoi{10.1086/509629}

\bibitem[{{Hubeny} {et~al.}(2001){Hubeny}, {Blaes}, {Krolik}, \&
  {Agol}}]{2001ApJ...559..680H}
{Hubeny}, I., {Blaes}, O., {Krolik}, J.~H., \& {Agol}, E. 2001, \apj, 559, 680,
  \dodoi{10.1086/322344}

\bibitem[{{Hunter}(2007)}]{Hunter2007}
{Hunter}, J.~D. 2007, Computing in Science and Engineering, 9, 90,
  \dodoi{10.1109/MCSE.2007.55}

\bibitem[{{Jalan} {et~al.}(2023){Jalan}, {Rakshit}, {Woo}, {Kotilainen}, \&
  {Stalin}}]{2023MNRAS.521L..11J}
{Jalan}, P., {Rakshit}, S., {Woo}, J.-J., {Kotilainen}, J., \& {Stalin}, C.~S.
  2023, \mnras, 521, L11, \dodoi{10.1093/mnrasl/slad014}

\bibitem[{{Jin} {et~al.}(2012){Jin}, {Ward}, \& {Done}}]{2012MNRAS.425..907J}
{Jin}, C., {Ward}, M., \& {Done}, C. 2012, \mnras, 425, 907,
  \dodoi{10.1111/j.1365-2966.2012.21272.x}

\bibitem[{{Kim} {et~al.}(2023){Kim}, {Im}, {Kim}, {Kim}, {Shin}, {Shim}, \&
  {Song}}]{2023ApJ...954..156K}
{Kim}, D., {Im}, M., {Kim}, M., {et~al.} 2023, \apj, 954, 156,
  \dodoi{10.3847/1538-4357/aceb5e}

\bibitem[{Kohonen(1982)}]{KOHONEN82}
Kohonen, T. 1982, Biological Cybernetics, 43, 59, \dodoi{10.1007/BF00337288}

\bibitem[{{Krawczyk} {et~al.}(2013){Krawczyk}, {Richards}, {Mehta}, {Vogeley},
  {Gallagher}, {Leighly}, {Ross}, \& {Schneider}}]{2013ApJS..206....4K}
{Krawczyk}, C.~M., {Richards}, G.~T., {Mehta}, S.~S., {et~al.} 2013, \apjs,
  206, 4, \dodoi{10.1088/0067-0049/206/1/4}

\bibitem[{{La Torre} \& {Pacucci}(2024)}]{2024LaTorre_Pacucci}
{La Torre}, V., \& {Pacucci}, F. 2024, arXiv e-prints, arXiv:2410.11951,
  \dodoi{10.48550/arXiv.2410.11951}

\bibitem[{{La Torre} {et~al.}(2024){La Torre}, {Sajina}, {Goulding},
  {Marchesini}, {Bezanson}, {Pearl}, \& {Sodr{\'e}}}]{2024LaTorre}
{La Torre}, V., {Sajina}, A., {Goulding}, A.~D., {et~al.} 2024, \aj, 167, 261,
  \dodoi{10.3847/1538-3881/ad3821}

\bibitem[{{Lawrence} {et~al.}(2007){Lawrence}, {Warren}, {Almaini}, {Edge},
  {Hambly}, {Jameson}, {Lucas}, {Casali}, {Adamson}, {Dye}, {Emerson},
  {Foucaud}, {Hewett}, {Hirst}, {Hodgkin}, {Irwin}, {Lodieu}, {McMahon},
  {Simpson}, {Smail}, {Mortlock}, \& {Folger}}]{2007MNRAS.379.1599L}
{Lawrence}, A., {Warren}, S.~J., {Almaini}, O., {et~al.} 2007, \mnras, 379,
  1599, \dodoi{10.1111/j.1365-2966.2007.12040.x}

\bibitem[{{Lusso} {et~al.}(2012){Lusso}, {Comastri}, {Simmons}, {Mignoli},
  {Zamorani}, {Vignali}, {Brusa}, {Shankar}, {Lutz}, {Trump}, {Maiolino},
  {Gilli}, {Bolzonella}, {Puccetti}, {Salvato}, {Impey}, {Civano}, {Elvis},
  {Mainieri}, {Silverman}, {Koekemoer}, {Bongiorno}, {Merloni}, {Berta}, {Le
  Floc'h}, {Magnelli}, {Pozzi}, \& {Riguccini}}]{2012MNRAS.425..623L}
{Lusso}, E., {Comastri}, A., {Simmons}, B.~D., {et~al.} 2012, \mnras, 425, 623,
  \dodoi{10.1111/j.1365-2966.2012.21513.x}

\bibitem[{{Lyke} {et~al.}(2020){Lyke}, {Higley}, {McLane}, {Schurhammer},
  {Myers}, {Ross}, {Dawson}, {Chabanier}, {Martini}, {Busca}, {Mas des
  Bourboux}, {Salvato}, {Streblyanska}, {Zarrouk}, {Burtin}, {Anderson},
  {Bautista}, {Bizyaev}, {Brandt}, {Brinkmann}, {Brownstein}, {Comparat},
  {Green}, {de la Macorra}, {Mu{\~n}oz Guti{\'e}rrez}, {Hou}, {Newman},
  {Palanque-Delabrouille}, {P{\^a}ris}, {Percival}, {Petitjean}, {Rich},
  {Rossi}, {Schneider}, {Smith}, {Vivek}, \& {Weaver}}]{2020ApJS..250....8L}
{Lyke}, B.~W., {Higley}, A.~N., {McLane}, J.~N., {et~al.} 2020, \apjs, 250, 8,
  \dodoi{10.3847/1538-4365/aba623}

\bibitem[{{Lynds}(1971)}]{Lynds1971}
{Lynds}, R. 1971, \apjl, 164, L73, \dodoi{10.1086/180695}

\bibitem[{{Lyu} \& {Rieke}(2017)}]{Lyu2017}
{Lyu}, J., \& {Rieke}, G.~H. 2017, \apj, 841, 76,
  \dodoi{10.3847/1538-4357/aa7051}

\bibitem[{{Marconi} {et~al.}(2004){Marconi}, {Risaliti}, {Gilli}, {Hunt},
  {Maiolino}, \& {Salvati}}]{2004MNRAS.351..169M}
{Marconi}, A., {Risaliti}, G., {Gilli}, R., {et~al.} 2004, \mnras, 351, 169,
  \dodoi{10.1111/j.1365-2966.2004.07765.x}

\bibitem[{{Martin} {et~al.}(2005){Martin}, {Fanson}, {Schiminovich},
  {Morrissey}, {Friedman}, {Barlow}, {Conrow}, {Grange}, {Jelinsky},
  {Milliard}, {Siegmund}, {Bianchi}, {Byun}, {Donas}, {Forster}, {Heckman},
  {Lee}, {Madore}, {Malina}, {Neff}, {Rich}, {Small}, {Surber}, {Szalay},
  {Welsh}, \& {Wyder}}]{2005ApJ...619L...1M}
{Martin}, D.~C., {Fanson}, J., {Schiminovich}, D., {et~al.} 2005, \apjl, 619,
  L1, \dodoi{10.1086/426387}

\bibitem[{{McMahon} {et~al.}(2013){McMahon}, {Banerji}, {Gonzalez}, {Koposov},
  {Bejar}, {Lodieu}, {Rebolo}, \& {VHS Collaboration}}]{2013McMahon}
{McMahon}, R.~G., {Banerji}, M., {Gonzalez}, E., {et~al.} 2013, The Messenger,
  154, 35

\bibitem[{{Merloni} {et~al.}(2024){Merloni}, {Lamer}, {Liu}, {Ramos-Ceja},
  {Brunner}, {Bulbul}, {Dennerl}, {Doroshenko}, {Freyberg}, {Friedrich},
  {Gatuzz}, {Georgakakis}, {Haberl}, {Igo}, {Kreykenbohm}, {Liu}, {Maitra},
  {Malyali}, {Mayer}, {Nandra}, {Predehl}, {Robrade}, {Salvato}, {Sanders},
  {Stewart}, {Tub{\'\i}n-Arenas}, {Weber}, {Wilms}, {Arcodia}, {Artis},
  {Aschersleben}, {Avakyan}, {Aydar}, {Bahar}, {Balzer}, {Becker}, {Berger},
  {Boller}, {Bornemann}, {Br{\"u}ggen}, {Brusa}, {Buchner}, {Burwitz},
  {Camilloni}, {Clerc}, {Comparat}, {Coutinho}, {Czesla}, {Dannhauer},
  {Dauner}, {Dauser}, {Dietl}, {Dolag}, {Dwelly}, {Egg}, {Ehl}, {Freund},
  {Friedrich}, {Gaida}, {Garrel}, {Ghirardini}, {Gokus}, {Gr{\"u}nwald},
  {Grandis}, {Grotova}, {Gruen}, {Gueguen}, {H{\"a}mmerich}, {Hamaus},
  {Hasinger}, {Haubner}, {Homan}, {Ider Chitham}, {Joseph}, {Joyce},
  {K{\"o}nig}, {Kaltenbrunner}, {Khokhriakova}, {Kink}, {Kirsch}, {Kluge},
  {Knies}, {Krippendorf}, {Krumpe}, {Kurpas}, {Li}, {Liu}, {Locatelli},
  {Lorenz}, {M{\"u}ller}, {Magaudda}, {Mannes}, {McCall}, {Meidinger},
  {Michailidis}, {Migkas}, {Mu{\~n}oz-Giraldo}, {Musiimenta}, {Nguyen-Dang},
  {Ni}, {Olechowska}, {Ota}, {Pacaud}, {Pasini}, {Perinati}, {Pires},
  {Pommranz}, {Ponti}, {Poppenhaeger}, {P{\"u}hlhofer}, {Rau}, {Reh},
  {Reiprich}, {Roster}, {Saeedi}, {Santangelo}, {Sasaki}, {Schmitt},
  {Schneider}, {Schrabback}, {Schuster}, {Schwope}, {Seppi}, {Serim},
  {Shreeram}, {Sokolova-Lapa}, {Starck}, {Stelzer}, {Stierhof}, {Suleimanov},
  {Tenzer}, {Traulsen}, {Tr{\"u}mper}, {Tsuge}, {Urrutia}, {Veronica},
  {Waddell}, {Willer}, {Wolf}, {Yeung}, {Zainab}, {Zangrandi}, {Zhang},
  {Zhang}, \& {Zheng}}]{2024A&A...682A..34M}
{Merloni}, A., {Lamer}, G., {Liu}, T., {et~al.} 2024, \aap, 682, A34,
  \dodoi{10.1051/0004-6361/202347165}

\bibitem[{Moosavi {et~al.}(2014)Moosavi, Packmann, \&
  Vall{\'e}s}]{moosavi2014sompy}
Moosavi, V., Packmann, S., \& Vall{\'e}s, I. 2014, SOMPY: A Python Library for
  Self Organizing Map (SOM)

\bibitem[{{Morrissey} {et~al.}(2007){Morrissey}, {Conrow}, {Barlow}, {Small},
  {Seibert}, {Wyder}, {Budav{\'a}ri}, {Arnouts}, {Friedman}, {Forster},
  {Martin}, {Neff}, {Schiminovich}, {Bianchi}, {Donas}, {Heckman}, {Lee},
  {Madore}, {Milliard}, {Rich}, {Szalay}, {Welsh}, \&
  {Yi}}]{2007ApJS..173..682M}
{Morrissey}, P., {Conrow}, T., {Barlow}, T.~A., {et~al.} 2007, \apjs, 173, 682,
  \dodoi{10.1086/520512}

\bibitem[{{Nemmen} \& {Brotherton}(2010)}]{2010Nemmen}
{Nemmen}, R.~S., \& {Brotherton}, M.~S. 2010, \mnras, 408, 1598,
  \dodoi{10.1111/j.1365-2966.2010.17224.x}

\bibitem[{{Novikov} \& {Thorne}(1973)}]{NT1973}
{Novikov}, I.~D., \& {Thorne}, K.~S. 1973, in Black Holes (Les Astres Occlus),
  343--450

\bibitem[{{Pedregosa} {et~al.}(2011){Pedregosa}, {Varoquaux}, {Gramfort},
  {Michel}, {Thirion}, {Grisel}, {Blondel}, {M{\"u}ller}, {Nothman}, {Louppe},
  {Prettenhofer}, {Weiss}, {Dubourg}, {Vanderplas}, {Passos}, {Cournapeau},
  {Brucher}, {Perrot}, \& {Duchesnay}}]{Pedregosa2011}
{Pedregosa}, F., {Varoquaux}, G., {Gramfort}, A., {et~al.} 2011, Journal of
  Machine Learning Research, 12, 2825, \dodoi{10.48550/arXiv.1201.0490}

\bibitem[{{Perez} \& {Granger}(2007)}]{Perez2007}
{Perez}, F., \& {Granger}, B.~E. 2007, Computing in Science and Engineering, 9,
  21, \dodoi{10.1109/MCSE.2007.53}

\bibitem[{{Prochaska} {et~al.}(2009){Prochaska}, {Worseck}, \&
  {O'Meara}}]{2009ApJ...705L.113P}
{Prochaska}, J.~X., {Worseck}, G., \& {O'Meara}, J.~M. 2009, \apjl, 705, L113,
  \dodoi{10.1088/0004-637X/705/2/L113}

\bibitem[{{Ranalli} {et~al.}(2015){Ranalli}, {Georgantopoulos}, {Corral},
  {Koutoulidis}, {Rovilos}, {Carrera}, {Akylas}, {Del Moro}, {Georgakakis},
  {Gilli}, \& {Vignali}}]{2015A&A...577A.121R}
{Ranalli}, P., {Georgantopoulos}, I., {Corral}, A., {et~al.} 2015, \aap, 577,
  A121, \dodoi{10.1051/0004-6361/201425246}

\bibitem[{{Ren} {et~al.}(2024){Ren}, {Guo}, {Shen}, {Silverman}, {Burke},
  {Wang}, \& {Wang}}]{2024arXiv240617598R}
{Ren}, W., {Guo}, H., {Shen}, Y., {et~al.} 2024, arXiv e-prints,
  arXiv:2406.17598, \dodoi{10.48550/arXiv.2406.17598}

\bibitem[{{Richards} {et~al.}(2006){Richards}, {Lacy}, {Storrie-Lombardi},
  {Hall}, {Gallagher}, {Hines}, {Fan}, {Papovich}, {Vanden Berk}, {Trammell},
  {Schneider}, {Vestergaard}, {York}, {Jester}, {Anderson}, {Budav{\'a}ri}, \&
  {Szalay}}]{2006ApJS..166..470R}
{Richards}, G.~T., {Lacy}, M., {Storrie-Lombardi}, L.~J., {et~al.} 2006, \apjs,
  166, 470, \dodoi{10.1086/506525}

\bibitem[{{Rosario} {et~al.}(2013){Rosario}, {Trakhtenbrot}, {Lutz}, {Netzer},
  {Trump}, {Silverman}, {Schramm}, {Lusso}, {Berta}, {Bongiorno}, {Brusa},
  {F{\"o}rster-Schreiber}, {Genzel}, {Lilly}, {Magnelli}, {Mainieri},
  {Maiolino}, {Merloni}, {Mignoli}, {Nordon}, {Popesso}, {Salvato}, {Santini},
  {Tacconi}, \& {Zamorani}}]{Rosario2013}
{Rosario}, D.~J., {Trakhtenbrot}, B., {Lutz}, D., {et~al.} 2013, \aap, 560,
  A72, \dodoi{10.1051/0004-6361/201322196}

\bibitem[{{Runnoe} {et~al.}(2012{\natexlab{a}}){Runnoe}, {Brotherton}, \&
  {Shang}}]{2012MNRAS.422..478R}
{Runnoe}, J.~C., {Brotherton}, M.~S., \& {Shang}, Z. 2012{\natexlab{a}},
  \mnras, 422, 478, \dodoi{10.1111/j.1365-2966.2012.20620.x}

\bibitem[{{Runnoe} {et~al.}(2012{\natexlab{b}}){Runnoe}, {Brotherton}, \&
  {Shang}}]{2012MNRAS.426.2677R}
---. 2012{\natexlab{b}}, \mnras, 426, 2677,
  \dodoi{10.1111/j.1365-2966.2012.21644.x}

\bibitem[{{Schlafly} \& {Finkbeiner}(2011)}]{2011ApJ...737..103S}
{Schlafly}, E.~F., \& {Finkbeiner}, D.~P. 2011, \apj, 737, 103,
  \dodoi{10.1088/0004-637X/737/2/103}

\bibitem[{{Schlafly} {et~al.}(2019){Schlafly}, {Meisner}, \&
  {Green}}]{Schlafly2019}
{Schlafly}, E.~F., {Meisner}, A.~M., \& {Green}, G.~M. 2019, \apjs, 240, 30,
  \dodoi{10.3847/1538-4365/aafbea}

\bibitem[{{Schlegel} {et~al.}(1998){Schlegel}, {Finkbeiner}, \&
  {Davis}}]{1998ApJ...500..525S}
{Schlegel}, D.~J., {Finkbeiner}, D.~P., \& {Davis}, M. 1998, \apj, 500, 525,
  \dodoi{10.1086/305772}

\bibitem[{{Schmidt} \& {Green}(1983)}]{1983ApJ...269..352S}
{Schmidt}, M., \& {Green}, R.~F. 1983, \apj, 269, 352, \dodoi{10.1086/161048}

\bibitem[{{Shakura} \& {Sunyaev}(1976)}]{SS1976}
{Shakura}, N.~I., \& {Sunyaev}, R.~A. 1976, \mnras, 175, 613,
  \dodoi{10.1093/mnras/175.3.613}

\bibitem[{{Shang} {et~al.}(2011){Shang}, {Brotherton}, {Wills}, {Wills},
  {Cales}, {Dale}, {Green}, {Runnoe}, {Nemmen}, {Gallagher}, {Ganguly},
  {Hines}, {Kelly}, {Kriss}, {Li}, {Tang}, \& {Xie}}]{2011Shang}
{Shang}, Z., {Brotherton}, M.~S., {Wills}, B.~J., {et~al.} 2011, \apjs, 196, 2,
  \dodoi{10.1088/0067-0049/196/1/2}

\bibitem[{{Shangguan} {et~al.}(2020){Shangguan}, {Ho}, {Bauer}, {Wang}, \&
  {Treister}}]{Shangguan2020}
{Shangguan}, J., {Ho}, L.~C., {Bauer}, F.~E., {Wang}, R., \& {Treister}, E.
  2020, \apj, 899, 112, \dodoi{10.3847/1538-4357/aba8a1}

\bibitem[{{Shen} {et~al.}(2011){Shen}, {Richards}, {Strauss}, {Hall},
  {Schneider}, {Snedden}, {Bizyaev}, {Brewington}, {Malanushenko},
  {Malanushenko}, {Oravetz}, {Pan}, \& {Simmons}}]{2011ApJS..194...45S}
{Shen}, Y., {Richards}, G.~T., {Strauss}, M.~A., {et~al.} 2011, \apjs, 194, 45,
  \dodoi{10.1088/0067-0049/194/2/45}

\bibitem[{{Sobolewska} {et~al.}(2004{\natexlab{a}}){Sobolewska},
  {Siemiginowska}, \& {{\.Z}ycki}}]{2004ApJ...608...80S}
{Sobolewska}, M.~A., {Siemiginowska}, A., \& {{\.Z}ycki}, P.~T.
  2004{\natexlab{a}}, \apj, 608, 80, \dodoi{10.1086/392529}

\bibitem[{{Sobolewska} {et~al.}(2004{\natexlab{b}}){Sobolewska},
  {Siemiginowska}, \& {{\.Z}ycki}}]{2004ApJ...617..102S}
---. 2004{\natexlab{b}}, \apj, 617, 102, \dodoi{10.1086/425262}

\bibitem[{{Stalevski} {et~al.}(2016){Stalevski}, {Ricci}, {Ueda}, {Lira},
  {Fritz}, \& {Baes}}]{2016MNRAS.458.2288S}
{Stalevski}, M., {Ricci}, C., {Ueda}, Y., {et~al.} 2016, \mnras, 458, 2288,
  \dodoi{10.1093/mnras/stw444}

\bibitem[{{Steffen} {et~al.}(2006){Steffen}, {Strateva}, {Brandt}, {Alexander},
  {Koekemoer}, {Lehmer}, {Schneider}, \& {Vignali}}]{2006AJ....131.2826S}
{Steffen}, A.~T., {Strateva}, I., {Brandt}, W.~N., {et~al.} 2006, \aj, 131,
  2826, \dodoi{10.1086/503627}

\bibitem[{{Su} {et~al.}(2025){Su}, {Guo}, {Qiao}, {Pei}, {Ho}, \&
  {Lacey}}]{Su2025}
{Su}, T., {Guo}, Q., {Qiao}, E., {et~al.} 2025, arXiv e-prints,
  arXiv:2501.10793, \dodoi{10.48550/arXiv.2501.10793}

\bibitem[{{Trammell} {et~al.}(2007){Trammell}, {Vanden Berk}, {Schneider},
  {Richards}, {Hall}, {Anderson}, \& {Brinkmann}}]{2007Trammell}
{Trammell}, G.~B., {Vanden Berk}, D.~E., {Schneider}, D.~P., {et~al.} 2007,
  \aj, 133, 1780, \dodoi{10.1086/511817}

\bibitem[{{Traulsen} {et~al.}(2020){Traulsen}, {Schwope}, {Lamer}, {Ballet},
  {Carrera}, {Ceballos}, {Coriat}, {Freyberg}, {Koliopanos}, {Kurpas},
  {Michel}, {Motch}, {Page}, {Watson}, \& {Webb}}]{2020A&A...641A.137T}
{Traulsen}, I., {Schwope}, A.~D., {Lamer}, G., {et~al.} 2020, \aap, 641, A137,
  \dodoi{10.1051/0004-6361/202037706}

\bibitem[{Troyanskaya {et~al.}(2001)Troyanskaya, Cantor, Sherlock, Brown,
  Hastie, Tibshirani, Botstein, \& Altman}]{Troyanskaya}
Troyanskaya, O., Cantor, M., Sherlock, G., {et~al.} 2001, Bioinformatics, 17,
  520, \dodoi{10.1093/bioinformatics/17.6.520}

\bibitem[{{Ueda} {et~al.}(2003){Ueda}, {Akiyama}, {Ohta}, \&
  {Miyaji}}]{2003ApJ...598..886U}
{Ueda}, Y., {Akiyama}, M., {Ohta}, K., \& {Miyaji}, T. 2003, \apj, 598, 886,
  \dodoi{10.1086/378940}

\bibitem[{{van der Walt} {et~al.}(2011){van der Walt}, {Colbert}, \&
  {Varoquaux}}]{scipy}
{van der Walt}, S., {Colbert}, S.~C., \& {Varoquaux}, G. 2011, Computing in
  Science and Engineering, 13, 22, \dodoi{10.1109/MCSE.2011.37}

\bibitem[{{Vanden Berk} {et~al.}(2001){Vanden Berk}, {Richards}, {Bauer},
  {Strauss}, {Schneider}, {Heckman}, {York}, {Hall}, {Fan}, {Knapp},
  {Anderson}, {Annis}, {Bahcall}, {Bernardi}, {Briggs}, {Brinkmann}, {Brunner},
  {Burles}, {Carey}, {Castander}, {Connolly}, {Crocker}, {Csabai}, {Doi},
  {Finkbeiner}, {Friedman}, {Frieman}, {Fukugita}, {Gunn}, {Hennessy},
  {Ivezi{\'c}}, {Kent}, {Kunszt}, {Lamb}, {Leger}, {Long}, {Loveday}, {Lupton},
  {Meiksin}, {Merelli}, {Munn}, {Newberg}, {Newcomb}, {Nichol}, {Owen}, {Pier},
  {Pope}, {Rockosi}, {Schlegel}, {Siegmund}, {Smee}, {Snir}, {Stoughton},
  {Stubbs}, {SubbaRao}, {Szalay}, {Szokoly}, {Tremonti}, {Uomoto}, {Waddell},
  {Yanny}, \& {Zheng}}]{2001AJ....122..549V}
{Vanden Berk}, D.~E., {Richards}, G.~T., {Bauer}, A., {et~al.} 2001, \aj, 122,
  549, \dodoi{10.1086/321167}

\bibitem[{VanderPlas {et~al.}(2012)VanderPlas, Connolly, Ivezić, \&
  Gray}]{VanderPlas2012}
VanderPlas, J., Connolly, A.~J., Ivezić, Z., \& Gray, A. 2012, in 2012
  Conference on Intelligent Data Understanding, 47--54,
  \dodoi{10.1109/CIDU.2012.6382200}

\bibitem[{{Vasudevan} \& {Fabian}(2007)}]{2007MNRAS.381.1235V}
{Vasudevan}, R.~V., \& {Fabian}, A.~C. 2007, \mnras, 381, 1235,
  \dodoi{10.1111/j.1365-2966.2007.12328.x}

\bibitem[{{Volonteri} {et~al.}(2003){Volonteri}, {Haardt}, \&
  {Madau}}]{2003ApJ...582..559V}
{Volonteri}, M., {Haardt}, F., \& {Madau}, P. 2003, \apj, 582, 559,
  \dodoi{10.1086/344675}

\bibitem[{{Wall} {et~al.}(2023){Wall}, {Kilic}, {Bergeron}, \&
  {Leiphart}}]{2023MNRAS.523.4067W}
{Wall}, R.~E., {Kilic}, M., {Bergeron}, P., \& {Leiphart}, N.~D. 2023, \mnras,
  523, 4067, \dodoi{10.1093/mnras/stad1699}

\bibitem[{{Wall} {et~al.}(2019){Wall}, {Kilic}, {Bergeron}, {Rolland},
  {Genest-Beaulieu}, \& {Gianninas}}]{2019MNRAS.489.5046W}
{Wall}, R.~E., {Kilic}, M., {Bergeron}, P., {et~al.} 2019, \mnras, 489, 5046,
  \dodoi{10.1093/mnras/stz2506}

\bibitem[{{Webb} {et~al.}(2020){Webb}, {Coriat}, {Traulsen}, {Ballet}, {Motch},
  {Carrera}, {Koliopanos}, {Authier}, {de la Calle}, {Ceballos}, {Colomo},
  {Chuard}, {Freyberg}, {Garcia}, {Kolehmainen}, {Lamer}, {Lin}, {Maggi},
  {Michel}, {Page}, {Page}, {Perea-Calderon}, {Pineau}, {Rodriguez}, {Rosen},
  {Santos Lleo}, {Saxton}, {Schwope}, {Tom{\'a}s}, {Watson}, \&
  {Zakardjian}}]{2020A&A...641A.136W}
{Webb}, N.~A., {Coriat}, M., {Traulsen}, I., {et~al.} 2020, \aap, 641, A136,
  \dodoi{10.1051/0004-6361/201937353}

\bibitem[{{Wright} {et~al.}(2010){Wright}, {Eisenhardt}, {Mainzer}, {Ressler},
  {Cutri}, {Jarrett}, {Kirkpatrick}, {Padgett}, {McMillan}, {Skrutskie},
  {Stanford}, {Cohen}, {Walker}, {Mather}, {Leisawitz}, {Gautier}, {McLean},
  {Benford}, {Lonsdale}, {Blain}, {Mendez}, {Irace}, {Duval}, {Liu}, {Royer},
  {Heinrichsen}, {Howard}, {Shannon}, {Kendall}, {Walsh}, {Larsen}, {Cardon},
  {Schick}, {Schwalm}, {Abid}, {Fabinsky}, {Naes}, \&
  {Tsai}}]{2010AJ....140.1868W}
{Wright}, E.~L., {Eisenhardt}, P. R.~M., {Mainzer}, A.~K., {et~al.} 2010, \aj,
  140, 1868, \dodoi{10.1088/0004-6256/140/6/1868}

\bibitem[{{Wu} \& {Shen}(2022)}]{2022ApJS..263...42W}
{Wu}, Q., \& {Shen}, Y. 2022, \apjs, 263, 42, \dodoi{10.3847/1538-4365/ac9ead}

\bibitem[{{York} {et~al.}(2000){York}, {Adelman}, {Anderson}, {Anderson},
  {Annis}, {Bahcall}, {Bakken}, {Barkhouser}, {Bastian}, {Berman}, {Boroski},
  {Bracker}, {Briegel}, {Briggs}, {Brinkmann}, {Brunner}, {Burles}, {Carey},
  {Carr}, {Castander}, {Chen}, {Colestock}, {Connolly}, {Crocker}, {Csabai},
  {Czarapata}, {Davis}, {Doi}, {Dombeck}, {Eisenstein}, {Ellman}, {Elms},
  {Evans}, {Fan}, {Federwitz}, {Fiscelli}, {Friedman}, {Frieman}, {Fukugita},
  {Gillespie}, {Gunn}, {Gurbani}, {de Haas}, {Haldeman}, {Harris}, {Hayes},
  {Heckman}, {Hennessy}, {Hindsley}, {Holm}, {Holmgren}, {Huang}, {Hull},
  {Husby}, {Ichikawa}, {Ichikawa}, {Ivezi{\'c}}, {Kent}, {Kim}, {Kinney},
  {Klaene}, {Kleinman}, {Kleinman}, {Knapp}, {Korienek}, {Kron}, {Kunszt},
  {Lamb}, {Lee}, {Leger}, {Limmongkol}, {Lindenmeyer}, {Long}, {Loomis},
  {Loveday}, {Lucinio}, {Lupton}, {MacKinnon}, {Mannery}, {Mantsch}, {Margon},
  {McGehee}, {McKay}, {Meiksin}, {Merelli}, {Monet}, {Munn}, {Narayanan},
  {Nash}, {Neilsen}, {Neswold}, {Newberg}, {Nichol}, {Nicinski}, {Nonino},
  {Okada}, {Okamura}, {Ostriker}, {Owen}, {Pauls}, {Peoples}, {Peterson},
  {Petravick}, {Pier}, {Pope}, {Pordes}, {Prosapio}, {Rechenmacher}, {Quinn},
  {Richards}, {Richmond}, {Rivetta}, {Rockosi}, {Ruthmansdorfer}, {Sandford},
  {Schlegel}, {Schneider}, {Sekiguchi}, {Sergey}, {Shimasaku}, {Siegmund},
  {Smee}, {Smith}, {Snedden}, {Stone}, {Stoughton}, {Strauss}, {Stubbs},
  {SubbaRao}, {Szalay}, {Szapudi}, {Szokoly}, {Thakar}, {Tremonti}, {Tucker},
  {Uomoto}, {Vanden Berk}, {Vogeley}, {Waddell}, {Wang}, {Watanabe},
  {Weinberg}, {Yanny}, {Yasuda}, \& {SDSS Collaboration}}]{York2000}
{York}, D.~G., {Adelman}, J., {Anderson}, Jr., J.~E., {et~al.} 2000, \aj, 120,
  1579, \dodoi{10.1086/301513}

\bibitem[{{Yuan} \& {Narayan}(2014)}]{2014ARA&A..52..529Y}
{Yuan}, F., \& {Narayan}, R. 2014, \araa, 52, 529,
  \dodoi{10.1146/annurev-astro-082812-141003}

\bibitem[{{Yuan} {et~al.}(2013){Yuan}, {Liu}, \& {Xiang}}]{2013MNRAS.430.2188Y}
{Yuan}, H.~B., {Liu}, X.~W., \& {Xiang}, M.~S. 2013, \mnras, 430, 2188,
  \dodoi{10.1093/mnras/stt039}

\bibitem[{{Zhuang} {et~al.}(2021){Zhuang}, {Ho}, \& {Shangguan}}]{Zhuang2021}
{Zhuang}, M.-Y., {Ho}, L.~C., \& {Shangguan}, J. 2021, \apj, 906, 38,
  \dodoi{10.3847/1538-4357/abc94d}

\end{thebibliography}
\bibliographystyle{aasjournal}



\end{document}